\documentclass[12pt]{article}

\input{epsf}

\usepackage{graphics}
\usepackage{subfigure}
\usepackage{setspace}
\usepackage{Tabbing}
\usepackage{extramarks}
\usepackage{chngpage}
\usepackage[usenames,dvipsnames]{color}
\usepackage{graphicx,float,wrapfig}
\usepackage{sidecap}
\usepackage[colorlinks,hyperindex]{hyperref}
\hypersetup{citecolor=Magenta, linkcolor=Blue, urlcolor=Red}

\def\rf#1{(\ref{eq:#1})}
\def\lab#1{\label{eq:#1}}

\def\br{\begin{eqnarray}}
\def\er{\end{eqnarray}}
\def\be{\begin{equation}}
\def\ee{\end{equation}}
\def\({\left(}
\def\){\right)}

\def\rlx{\relax\leavevmode}
\def\IR{\rlx\hbox{\rm I\kern-.18em R}}

\def\u2{\mid u\mid^2}
\newcommand{\sbr}[2]{\left\lbrack\,{#1}\, ,\,{#2}\,\right\rbrack}

\def\IZ{\rlx\hbox{\sf Z\kern-.4em Z}}
\def\IR{\rlx\hbox{\rm I\kern-.18em R}}
\def\IC{\rlx\hbox{\,$\inbar\kern-.3em{\rm C}$}}
\def\one{\hbox{{1}\kern-.25em\hbox{l}}}

\input{epsf}

\usepackage{graphics}

\begin{document}

\begin{titlepage}
\vspace*{-1cm}

\vskip 1cm

\vspace{.2in}
\begin{center}
{\large\bf The integral equations of Yang-Mills and its gauge invariant conserved charges}
\end{center}

\vspace{.5cm}

\begin{center}
  L. A. Ferreira\footnote{e-mail: {\tt laf@ifsc.usp.br}} and G. Luchini\footnote{e-mail: {\tt gabriel.luchini@gmail.com}}

\vspace{.5 in}
\small

\par \vskip .2in \noindent
 Instituto de F\'\i sica de S\~ao Carlos; IFSC/USP;\\
Universidade de S\~ao Paulo - USP \\ 
Caixa Postal 369, CEP 13560-970, S\~ao Carlos-SP, Brazil

\normalsize
\end{center}

\vspace{.2in}

\begin{abstract}
Despite the fact that the integral form of the equations of classical electrodynamics is well known, the same is not true for non-abelian gauge theories. The aim of the present paper is threefold. First, we present the integral form of the classical Yang-Mills equations in the presence of sources, and then use it to solve the long standing problem of constructing conserved charges, for any field configuration, which are invariant under general gauge transformations and not only under transformations that go to a constant at spatial infinity. The construction is based on concepts in loop spaces and on  a generalization of the non-abelian Stokes theorem for two-form connections.   The third goal of the paper is to present the integral form of the self dual Yangs-Mills equations, and calculate the conserved charges associated to them. The charges are explicitly evaluated for the cases of monopoles, dyons, instantons and merons, and we show that in many cases those charges must be quantized. Our  results  are important in the understanding  of global properties of non-abelian gauge theories.

\end{abstract}


\end{titlepage}

\section{Introduction}
\label{sec:intro}
\setcounter{equation}{0}

The integral form of the equations of  classical electrodynamics precedes Maxwell equations and play a crucial role in the understanding of electromagnetic phenomena. The non-abelian gauge theories have been originally formulated in a differential form through the Yang-Mills equations, and its integral formulation has not been constructed.  
The aim of the present paper is threefold. First we present the integral form of the classical equations of motion of non-abelian gauge theories, which allows us to present the Yang-Mills equations as the equality of an ordered volume integral to an ordered surface integral on its border. Our construction was made possible by the use of  a generalization of the non-abelian Stokes theorem  for two-form connections proposed some years ago  in the context of integrable field theories in dimensions higher than two \cite{afs1,afs2}.  The volume ordered integral present some highly non-trivial and non-linear terms, involving the field tensor and its Hodge dual, which certainly will play an important role in the global aspects of the Yang-Mills theory.  The differential Yang-Mills equations are recovered in the limit when the volume is taken to be infinitesimal. The second goal of the paper is to solve the long standing problem of the construction of conserved charges that are invariant under general gauge transformations. As it is well known, the conserved charges presented in the Yang and Mills original paper \cite{ym}, as well as in modern textbooks,  are invariant under gauge transformations where the group element, performing the transformation, goes to a constant at spatial infinity. In the last decades several attempts were made to find truly gauge invariant  conserved charges using several techniques \cite{attempts}. We show that our integral form of Yang-Mills equations becomes a conservation law when the volume where it is considered is a closed volume, i.e. a three dimensional sub-manifold of the four dimensional space-time which has no border. Using appropriate boundary conditions we obtain a closed expression for the conserved charges as the eigenvalues of an operator obtained by a volume ordered integral, over the entire spatial sub-manifold (fixed time). As a consequence of our integral Yang-Mills equations that operator can also be written as an ordered surface integral on the border of the spatial sub-manifold. That fact makes the evaluation of the charges much simpler. We then show that such charges are invariant under general gauge transformation, independent of the parameterization of the volume, and on the reference point used on that parameterization. The expression is valid for any field configuration, and we evaluate it for well known solutions like monopoles, dyons, instantons and merons. 

The third goal of the paper is to construct the integral form of the self dual Yang-Mills equations. That is obtained in a similar manner as that of the integral form of the full Yang-Mills equations, but using instead the usual non-abelian Stokes for one-form connections. We show that such integral formulation leads to gauge invariant conserved quantities, which are invariant under reparameterization of surfaces and independent of the reference point used in such parameterization. The novelty is that the charges are given by the eigenvalues of surface ordered integrals of the field tensor and its Hodge dual, and they are shown to be constant on the other two coordinates perpendicular to that surface. 

The examples of monopoles, dyons, instantons and merons discussed in the paper are very important to shed light on the physical properties of the conserved charges.  The first point is that for those well known solutions the surface ordering becomes irrelevant in the evaluation of the operators leading to the charges. In fact, those operators tend to lie in the center of the gauge group, and the physical charges are identified with the eigenvalues of the Lie algebra elements which lead to those group elements under the exponential map.  That relation between charges and elements in the center of the group leads, in many cases, to the quantization of the physical charges. Another point is that the charges are associated to the abelian subgroup of the gauge group. In the examples we have worked out there are no gauge invariant conserved charges associated to the generators lying outside the abelian subalgebra. A further point is that the charges associated to the Wu-Yang monopole and the 't Hooft-Polyakov monopole are identical. In addition, they are shown to be conserved due to the equations of motion and no topological arguments are used. The evaluation of the charges do not involve the Higgs field and seems not to pay attention  to the symmetry breaking pattern. The construction of the magnetic charges in our paper differs from the usual techniques used in the literature involving homotopy invariant quantities, even though it leads to the same results.  But our methods allow to evaluate charges for cases that were not really known in the literature. We show that the merons carry a charge, conserved in the euclidean time, which is identical to the magnetic charge of the  Wu-Yang and  't Hooft-Polyakov monopoles. 

Our construction explores loop space techniques used in the study of integrable theories in any dimension \cite{afs1,afs2}, and may be important in the understanding of the integrability properties of Yang-Mills theory as well as of its self-dual sector. In fact, the most appropriate mathematical language to phrase our results is that of generalized loop spaces. There is a quite vast literature on integral and loop space formulations of gauge theories \cite{loopgauge}. Our approach differs in many aspects of those formulations even though it shares some of the ideas and insights permeating them. We stress however that the relevant loop space in our formulation is that of the maps from the two-sphere $S^2$ (and not from the circle $S^1$) onto the space-time, in the case of the full Yang-Mills equations. For the self dual sector however, the relevant loop space is that of the maps from the circle $S^1$ onto the space-time. 

The connection between integrable field theories in any dimension and loop space techniques by exploring the integral form of the equations of motion has been studied in  \cite{second}. It was shown there that integrable field theories in $1+1$ dimensions, Chern-Simons theory in $2+1$ dimensions, and Yang-Mills theory in $2+1$ and $3+1$ dimensions, all admit a uniform formulation in terms of integral equations on loop spaces, leading to a general and unique method for constructing conserved charges. The integral form of the Yang-Mills equations in $3+1$, discussed in this paper, already appears in \cite{second}. In the present paper we discuss further the physical consequences for non-abelian gauge theories of the existence of such integral equation. In addition, we show that the self-dual sector of Yang-Mills also admit an  integral equation, and that it leads in a similar way to new conserved quantities. The second part of the present paper is dedicated to the application of our ideas to well known solutions of Yang-Mills theory, like Wu-Yang and 'tHooft-Polyakov monopoles and dyons, as well as Euclidean solutions like instantons and merons. That is a very important contribution of the paper since the conserved charges evaluated for those solutions are new and were not explored in the literature before. We believe that  most of the    physical consequences of those charges are still to be explored and perhaps we have to consider the quantum theory to fully understand them. 

The paper is organized as follows: in section \ref{sec:statements} we present the main results of the paper in the form of very precise statements. It is stated the integral equations for the full Yang-Mills theory as well as for its self-dual sector. In this section we also present the closed expressions for the conserved charges. In section   \ref{sec:proofselfdual} we give the proof for the integral equation for the self-dual sector of the Yang-Mills theory using the ordinary non-abelian Stokes theorem.  In section \ref{sec:stokes} we give the proof for the generalized non-abelian Stokes theorem for a two-form connection based on the results of \cite{afs1,afs2}. Then in section \ref{sec:proofym} we use that theorem to prove the integral equations for the full Yang-Mills theory. In section \ref{sec:charges} we discuss some consequences of our integral equations and give the detailed construction of the conserved charges for the full Yang-Mills equations as well as for its self-dual sector.  In section \ref{sec:examples} we discuss the examples of monopoles, dyons, instantons and merons, and explicitly evaluated the conserved charges for all those solutions. Finally in appendix \ref{sec:wilson} we show how to (classically) regularize the Wilson line operator in order to evaluate the charges of the Wu-Yang monopole and dyon solutions.

\section{The main statements}
\label{sec:statements}
\setcounter{equation}{0}

{\bf The integral Yang-Mills equation and its charges.} Consider a Yang-Mills theory  for a gauge group $G$, with gauge field $A_{\mu}$, in the presence of matter currents $J^{\mu}$, on a  four dimensional space-time $M$. Let $\Omega$ be any tridimensional (topologically trivial)  volume on $M$, and $\partial \Omega$ be its border. We choose a reference point $x_R$ on $\partial \Omega$ and scan $\Omega$ with closed surfaces, based on $x_R$,  labelled by $\zeta$, and we scan the closed surfaces with closed loops based on $x_R$, labelled by $\tau$, and parametrized by $\sigma$, as we describe below. The classical dynamics of the gauge fields is governed by the following integral equations, on any such volume $\Omega$, 
\be
P_2 e^{ie\int_{\partial\Omega}d\tau d\sigma \left[ \alpha F_{\mu\nu}^W+\beta {\widetilde F}_{\mu\nu}^W\right] \frac{dx^{\mu}}{d\sigma}\frac{dx^{\nu}}{d\tau}}=  P_3e^{\int_{\Omega} d\zeta d\tau  V{\cal J}V^{-1}}
\lab{basic}
\ee
where $P_2$ and $P_3$ means surface and volume ordered integration respectively, ${\widetilde F}_{\mu\nu}$ is the Hodge dual of the field tensor, i.e. 
\be
F_{\mu\nu}=\partial_{\mu}A_{\nu}-\partial_{\nu}A_{\mu}+i\,e\,\sbr{A_{\mu}}{A_{\nu}} \qquad \qquad \qquad
{\widetilde F}_{\mu\nu}\equiv \frac{1}{2}\,\varepsilon_{\mu\nu\rho\lambda}\, F^{\rho\lambda}
\lab{fieldtensor}
\ee
where  $e$ is the  gauge coupling constant, $\alpha$ and $\beta$ are free parameters,  and where we have used the notation $X^W\equiv W^{-1}\,X\,W$, with $W$ being the  Wilson line defined on a curve $\Gamma$, parameterized by $\sigma$, through the equation
\be
\frac{d\,W}{d\,\sigma}+   i\,e\,A_{\mu}\,\frac{d\,x^{\mu}}{d\,\sigma}\,W=0
\lab{eqforw}
\ee
where $x^{\mu}$ ($\mu=0,1,2,3$) are the coordinates on the  four dimensional space-time $M$. The quantity $V$ is defined on a surface $\Sigma$ through the equation
\be
\frac{d\,V}{d\,\tau}-V\, T\(A,\tau\)=0
\lab{eqforv}
\ee
with 
\be
T\(A,\tau\)\equiv ie\,
 \int_{0}^{2\pi}d\sigma W^{-1}\left[ \alpha F_{\mu\nu}+\beta{\widetilde F}_{\mu\nu}\right] W \frac{dx^{\mu}}{d\sigma}\frac{dx^{\nu}}{d\tau}
 \lab{tdef}
 \ee
and where 
\br
{\cal J}&\equiv&
\int_0^{2\pi}d\sigma\left\{ ie\beta {\widetilde J}_{\mu\nu\lambda}^W
\frac{dx^{\mu}}{d\sigma}\frac{dx^{\nu}}{d\tau}
\frac{dx^{\lambda}}{d\zeta} \right. \nonumber\\
&+& \left. e^2\int_0^{\sigma}d\sigma^{\prime}
\sbr{\(\(\alpha-1\) F_{\kappa\rho}^W+\beta {\widetilde F}_{\kappa\rho}^W\)\(\sigma^{\prime}\)}
{\(\alpha F_{\mu\nu}^W+\beta {\widetilde F}_{\mu\nu}^W\)\(\sigma\)} \right.
\nonumber\\
&&\left. \times
\, \frac{d\,x^{\kappa}}{d\,\sigma^{\prime}}\frac{d\,x^{\mu}}{d\,\sigma}
\(\frac{d\,x^{\rho}\(\sigma^{\prime}\)}{d\,\tau}\frac{d\,x^{\nu}\(\sigma\)}{d\,\zeta}
-\frac{d\,x^{\rho}\(\sigma^{\prime}\)}{d\,\zeta}\frac{d\,x^{\nu}\(\sigma\)}{d\,\tau}\)\right\}
\lab{caljdef}
\er  
where ${\widetilde J}_{\mu\nu\lambda}$ is the Hodge dual of the current, i.e. $J^{\mu}=\frac{1}{3!}\varepsilon^{\mu\nu\rho\lambda}\,{\widetilde J}_{\nu\rho\lambda}$. The Yang-Mills equations are recovered from \rf{basic} in the case where $\Omega$ is taken to be an infinitesimal volume. Under appropriate boundary conditions the conserved charges are the eigenvalues of the operator
\be
Q_S=P_2 e^{ie\int_{\partial S}d\tau d\sigma\, W^{-1}\, (\alpha F_{\mu\nu}+\beta {\widetilde F}_{\mu\nu}) \,W\,\frac{dx^{\mu}}{d\sigma}\frac{dx^{\nu}}{d\tau}}=  P_3e^{\int_{S} d\zeta d\tau  V{\cal J}V^{-1}}
\lab{charge}
\ee
where $S$ is the $3$-dimensional spatial sub-manifold of $M$. Equivalently the charges are ${\rm Tr} Q_S^N$.  

{\bf The integral  self-dual Yang-Mills equation  and its charges.} Consider the self-dual sector of the Yang-Mills theory  defined by the first order differential equations
\be
F_{\mu\nu}=\kappa\, {\widetilde F}_{\mu\nu}
\lab{selfdualym}
\ee
where   $\kappa$ are the eigenvalues of the Hodge dual operation.   We shall be interested here in the case of an Euclidean  space-time where $\kappa = \pm 1$.  We propose that the integral equation for the self-dual sector of the Yang-Mills theory is  given by
\be
P_1\,e^{-ie\oint_{\partial \Sigma}d\sigma\, A_{\mu}\frac{dx^{\mu}}{d\sigma}} = 
P_2\,e^{ie\int_{\Sigma}d\sigma d\tau\,W^{-1}\left[\alpha\,F_{\mu\nu}+\kappa\,\(1-\alpha\)\,{\widetilde F}_{\mu\nu}\right]W\frac{dx^{\mu}}{d\sigma}\frac{dx^{\nu}}{d\tau}}
\lab{integralselfdualym}
\ee
where $\Sigma$ is any two-dimensional surface in the space-time $M$,  $\partial\Sigma$ is its border,  and $\alpha$ is a free parameter. The symbols $P_1$ and $P_2$ mean path and surface ordered integration respectively, and those are performed as follows. We choose a reference point $x_R$ on the border of $\Sigma$, and scan $\Sigma$ with closed loops, labelled by $\tau$, starting and ending at $x_R$, such that $\tau =0$ corresponds to the infinitesimal loop around $x_R$, and $\tau=2\pi$ corresponds to the border $\partial \Sigma$. Each loop is parameterize by $\sigma$, such that $\sigma=0$ and $\sigma=2\pi$ correspond to $x_R$. The quantity $W$ appearing on the r.h.s. of \rf{integralselfdualym} is obtained by integrating \rf{eqforw} on each loop from $\sigma =0$ (i.e. $x_R$) up to the point of the loop corresponding to $\sigma$ where the integrand $\left[\alpha\,F_{\mu\nu}+\kappa\,\(1-\alpha\)\,{\widetilde F}_{\mu\nu}\right]$ is evaluated. The l.h.s. of \rf{integralselfdualym} is obtained by integrating \rf{eqforw}  along the border of $\Sigma$. On the other hand the r.h.s. of  \rf{integralselfdualym}  comes from the integration of \rf{eqforv} with the same $T\(A,\tau\)$ given in \rf{tdef}, but with $\beta$ replaced by $\kappa\,\(1-\alpha\)$. 

The self-dual equations \rf{selfdualym} are recovered from \rf{integralselfdualym} in the limit where the surface $\Sigma$ is taken to be infinitesimal. The conserved charges associated to \rf{integralselfdualym} are constructed as follows: consider any two-dimensional plane  ${\cal D}_{\infty}$ in the space-time $M$, and let  $S_{\infty}^1$ be its border, i.e. a circle of infinite radius on that plane. Under appropriate boundary conditions, the eigenvalues of the operator 
\be
V\({\cal D}_{\infty}\)  =
P_2\,e^{ie\int_{{\cal D}_{\infty}}d\sigma d\tau\,W^{-1}\left[\alpha\,F_{\mu\nu}+\kappa\,\(1-\alpha\)\,{\widetilde F}_{\mu\nu}\right]W\frac{dx^{\mu}}{d\sigma}\frac{dx^{\nu}}{d\tau}}
= P_1\,e^{-ie\oint_{S_{\infty}^1}d\sigma\, A_{\mu}\frac{dx^{\mu}}{d\sigma}}
\lab{chargeselfdual}
\ee
are constants, i.e. independent of the two coordinates associated to the two axis perpendicular to ${\cal D}_{\infty}$ in $M$. 

{\bf On the nature of the eigenvalues.} All the quantities appearing in the formulas above are either elements of the gauge Lie algebra (like $A_{\mu}$ and $F_{\mu\nu}$) or of the gauge Lie group (like $W$ and $V$).  In order to perform the calculations  however we need  to choose a matrix representation of the Lie group (or equivalently of the Lie algebra) because equations like \rf{eqforw} and \rf{eqforv} involve the product of Lie algebra and Lie group quantities and so a definite representation has to be used. However, the choice of that representation is quite arbitrary. Note that not even the representation under which the matter fields transform under gauge transformations is relevant. Indeed, in the case of fermions for instance the current has the form $J_{\mu} \sim \eta^{ab}\,{\bar \psi}_i \,\gamma_{\mu}\, R_{ij}\(T_a\)\, \psi_j\, T_b$, where $T_a$, $a=1,2,\dots {\rm dim}\, G$, are the generators of the gauge group $G$, $\eta^{ab}$ is the inverse of the Killing form of $G$, and $R_{ij}$ is the matrix representation under which the spinors $\psi_i$ transform,  $i,j=1,2,\ldots {\rm dim}\,R$. Therefore, $J_{\mu}\equiv J_{\mu}^b\, T_b$ is an element of the Lie algebra for any choice of $R$, and in order to perform our calculations $J_{\mu}$ can be taken in any matrix representation irrespective of $R$. Consequently, the conserved charges which are the eigenvalues of the operators \rf{charge} and \rf{chargeselfdual}, which are in fact Lie group elements,  correspond  to the eigenvalues   of those operators in the  chosen representation of $G$. That choice of representation however  is arbitrary. We face then two possibilities. The eigenvalues can be different in different representations and then one finds an infinite spectrum of conserved charges, or then there is only a finite number of charges and the eigenvalues of the operators \rf{charge} and \rf{chargeselfdual} are the same in large (infinite) classes of representations. As we will show in the examples of monopoles, dyons, instantons and merons, the second possibility happens, i.e. we find a finite number of charges, and the operators \rf{charge} and \rf{chargeselfdual} evaluated on those solutions tend to lie in the center of the gauge group. In fact, we show that  the path and surface orderings become irrelevant for those solutions and the operators \rf{charge} and \rf{chargeselfdual} are expressed as (products of) ordinary exponentials of Lie algebra elements. The physical interpretation of the charges turn out to be associated to the eigenvalues of those Lie algebra elements and not really to the eigenvalues of the group elements \rf{charge} and \rf{chargeselfdual}.  The connection between  the eigenvalues of the Lie group and Lie algebra element leads, in many cases, to the quantization of the physical charges.

\section{The construction of the integral equation for the self-dual sector}
\label{sec:proofselfdual}
\setcounter{equation}{0}

In order to prove that \rf{integralselfdualym} does correspond to the integral form of the self dual equations \rf{selfdualym} we use the ordinary non-abelian Stokes theorem for a one-form connection $C_{\mu}$ given by \cite{arefeva,afs1,afs2}
\be
P_1\,e^{-\oint_{\partial \Sigma}d\sigma\, C_{\mu}\frac{dx^{\mu}}{d\sigma}} \, W_R= 
W_R\,P_2\,e^{\int_{\Sigma}d\sigma d\tau\,W^{-1}\,H_{\mu\nu}\,W\frac{dx^{\mu}}{d\sigma}\frac{dx^{\nu}}{d\tau}}
\lab{usualstokes}
\ee
with $H_{\mu\nu}=\partial_{\mu}C_{\nu}-\partial_{\nu}C_{\mu}+\sbr{C_{\mu}}{C_{\nu}}$, being the curvature of the connection $C_{\mu}$, and where $W_R$ is an integration constant, the value of $W$ at $x_R$. The meanings of the path and surface ordered integrations are the same as that in \rf{integralselfdualym}. For a simple and concise proof of the theorem \rf{usualstokes} see section 2 of \cite{afs1}. The proof of the non-abelian Stokes theorem \rf{usualstokes} does not rely on the use of a metric tensor, and so it is valid on any space-time (flat or curved) of any dimension  with any metric. The only requirements are  that the surfaces $\Sigma$ are topologically trivial (no holes or handles) and that the connection is a regular function of the space-time coordinates. Note that one can  obtain \rf{integralselfdualym} from \rf{usualstokes} by the identifications
\be
C_{\mu}\equiv ie\, A_{\mu}\; ;\qquad\qquad H_{\mu\nu}\equiv i e\, \left[\alpha\,F_{\mu\nu}+\kappa\,\(1-\alpha\)\,{\widetilde F}_{\mu\nu}\right]\; ; \qquad \qquad W_R\in Z\(G\)
\lab{identify}
\ee
where $Z\(G\)$ is the centre of the gauge group $G$. 
However, the first equation above implies that
\be
H_{\mu\nu}= i e\, F_{\mu\nu}
\lab{gfrel}
\ee
The compatibility between  \rf{identify} and \rf{gfrel} is provided by \rf{selfdualym}. Note that the case $\alpha=1$ is trivial since it leads to an identity. Therefore, \rf{integralselfdualym} is a direct consequence of the non-abelian Stokes theorem \rf{usualstokes} and the self dual Yang-Mills equations \rf{selfdualym}. The condition that the integration constant $W_R$ has to belong to $Z\(G\)$ comes from the requirement that \rf{integralselfdualym} has to transform covariantly under gauge transformations. The argument for that is similar to the one used in the paragraph below \rf{gaugetransf}, in the context of the integral equation for the full Yang-Mills equations. 
 
 On the other hand the integral equations  \rf{integralselfdualym} imply the differential equations \rf{selfdualym} in the limit where the surface $\Sigma$ is infinitesimal. Indeed, take $\Sigma$ to be of  rectangular shape on the plane $x^{\mu}\,x^{\nu}$, with  infinitesimal sides $\delta x^{\mu}$ and $\delta x^{\nu}$ ($\mu$ and $\nu$ fixed). We then evaluate both sides of \rf{integralselfdualym} by Taylor expanding the integrands around one given corner of the rectangle and keeping things up to first non-trivial order. One can check that the l.h.s. of \rf{integralselfdualym} gives $\left[\one + ie\,F_{\mu\nu}\,\delta x^{\mu}\,\delta x^{\nu}\right]$, with no sum in $\mu$ and $\nu$. In addition, the r.h.s. of \rf{integralselfdualym} gives, up to first non-trivial order, $\left[\one + i e\, \(\alpha\,F_{\mu\nu}+\kappa\,\(1-\alpha\)\,{\widetilde F}_{\mu\nu}\)\,\delta x^{\mu}\,\delta x^{\nu}\right]$ (again no sum in $\mu$ and $\nu$). By equating those two quantities one obtains \rf{selfdualym}  for any value of $\alpha$, except $\alpha=1$, which should be excluded. 

\section{The generalized non-abelian Stokes theorem}
\label{sec:stokes}
\setcounter{equation}{0}

 In order to prove that \rf{basic} does correspond to an integral formulation of the classical Yang-Mills dynamics, we shall start by describing the generalization of the non-abelian Stokes theorem  as  formulated in \cite{afs1,afs2}. Consider a surface $\Sigma$ scanned by a set of closed loops with common base point $x_{R}$ on the border $\partial \Sigma$. The points on the loops are parameterized by  $\sigma \in [0,2\pi]$ and each loop is labeled by a parameter $\tau$ such that $\tau=0$ corresponds to the infinitesimal loop around $x_R$, and $\tau=2\pi$ to the border $\partial \Sigma$. We then introduce, on each point of $M$,  a rank two antisymmetric tensor $B_{\mu\nu}$ taking values on the Lie algebra ${\cal G}$ of $G$, and construct a quantity $V$ on the surface $\Sigma$ through   
\be
\frac{d\,V}{d\,\tau}-V\, T\(B,A,\tau\)=0 \qquad {\rm with} \qquad
T\(B,A,\tau\)\equiv
 \int_{0}^{2\,\pi}d\sigma\; W^{-1}\,B_{\mu\nu}\,W\, \frac{d\,x^{\mu}}{d\,\sigma}\,\frac{d\,x^{\nu}}{d\,\tau}
\lab{eqforvb}
\ee
 where the $\sigma$-integration is along the loop $\Gamma$ labeled by $\tau$, and $W$ is obtained from \rf{eqforw}, by integrating it along $\Gamma$ from the reference point $x_R$ to the point labeled by $\sigma$, where $B_{\mu\nu}$ is evaluated. By integrating \rf{eqforv}, from the infinitesimal loop around $x_R$ to the border of $\Sigma$, we obtain 
\be
V= V_R\; P_2 e^{\int_0^{2\,\pi}d\tau\int_0^{2\,\pi}d\sigma W^{-1} B_{\mu\nu}W \frac{dx^{\mu}}{d\sigma}\,\frac{dx^{\nu}}{d \tau}}
\lab{firstvint}
\ee 
where  $P_2$ means surface ordering according to the parameterization of $\Sigma$ as described above, and $V_R$ is an integration constant corresponding to the value of $V$ on an infinitesimal surface around $x_R$.
If one changes $\Sigma$, keeping its border fixed, by making variations $\delta x^{\mu}$ perpendicular to $\Sigma$ then $V$ varies according to 
\br
\delta V\,V^{-1}&\equiv&
\int_0^{2\,\pi}d\tau\,\int_0^{2\,\pi}d\sigma\,V\(\tau\)\,\left\{W^{-1}\,
\left[D_{\lambda}B_{\mu\nu}+D_{\mu}B_{\nu\lambda}+D_{\nu}B_{\lambda\mu}\right]
\,
W\frac{d\,x^{\mu}}{d\,\sigma}\,\frac{d\,x^{\nu}}{d\,\tau}\,
\delta x^{\lambda}\right. \nonumber\\
&-&\left. \int_0^{\sigma}d\sigma^{\prime}
\sbr{B_{\kappa\rho}^W\(\sigma^{\prime}\)-ie F_{\kappa\rho}^W\(\sigma^{\prime}\)}
{B_{\mu\nu}^W\(\sigma\)}\frac{dx^{\kappa}}{d\sigma^{\prime}}\frac{dx^{\mu}}{d\sigma}\right.\nonumber\\
&&\left.\times
\(\frac{d\,x^{\rho}\(\sigma^{\prime}\)}{d\,\tau}\delta x^{\nu}\(\sigma\)
-\delta x^{\rho}\(\sigma^{\prime}\)\,\frac{d\,x^{\nu}\(\sigma\)}{d\,\tau}\)\right\} V^{-1}\(\tau\)
\lab{deltav}
\er  
where  $D_{\mu}*=\partial_{\mu}*+i\,e\,\sbr{A_{\mu}}{*}$. For a detailed account on how to obtain \rf{deltav} see sec. 5.3 of \cite{afs1},  or sec. 2.3 of \cite{afs2}, or then the appendix of \cite{second}. The quantity $V\(\tau\)$ appearing on the r.h.s. of \rf{deltav} is obtained by integrating \rf{eqforvb} from the infinitesimal loop around $x_R$ to the  loop labelled by $\tau$ on the scanning of $\Sigma$ described above. Note that the two $\sigma$-integrations on the second term on the r.h.s. of \rf{deltav} are performed on the same loop labelled by $\tau$. Consider now the case where the surface $\Sigma$ is closed, and the border of $\Sigma$ is  contracted to $x_R$. The expression \rf{deltav} gives then the variation of $V$ when we vary $\Sigma$ keeping  $x_R$ fixed. Therefore, if one starts with an infinitesimal closed surface $\Sigma_{R}$ around $x_R$ one can blows it up until it becomes $\Sigma$. One can label all those closed surfaces using a parameter $\zeta \in [0,2\pi]$, such that $\zeta=0$ corresponds to $\Sigma_{R}$ and $\zeta=2\,\pi$ to $\Sigma$. The expression \rf{deltav} can be seen as a differential equation on $\zeta$ defining $V$ on the surface $\Sigma$, i.e.
\be
\frac{d\,V}{d\,\zeta} - {\cal K}\, V=0
\lab{eqforv2}
\ee
where ${\cal K}$ corresponds to the r.h.s. of  \rf{deltav} with $\delta x^{\mu}$ replaced by $\frac{d\,x^{\mu}}{d\,\zeta}$, i.e. 
\br
{\cal K}&\equiv&
\int_0^{2\,\pi}d\tau\,\int_0^{2\,\pi}d\sigma\,V\(\tau\)\,\left\{W^{-1}\,
\left[D_{\lambda}B_{\mu\nu}+D_{\mu}B_{\nu\lambda}+D_{\nu}B_{\lambda\mu}\right]
\,
W\frac{d\,x^{\mu}}{d\,\sigma}\,\frac{d\,x^{\nu}}{d\,\tau}\,
\delta x^{\lambda}\right. \nonumber\\
&-&\left. \int_0^{\sigma}d\sigma^{\prime}
\sbr{B_{\kappa\rho}^W\(\sigma^{\prime}\)-ie F_{\kappa\rho}^W\(\sigma^{\prime}\)}
{B_{\mu\nu}^W\(\sigma\)}\frac{dx^{\kappa}}{d\sigma^{\prime}}\frac{dx^{\mu}}{d\sigma}\right.\nonumber\\
&&\left.\times
\(\frac{d\,x^{\rho}\(\sigma^{\prime}\)}{d\,\tau}\frac{d\,x^{\nu}\(\sigma\)}{d\,\zeta}
-\frac{d\,x^{\rho}\(\sigma^{\prime}\)}{d\,\zeta}\,\frac{d\,x^{\nu}\(\sigma\)}{d\,\tau}\)\right\} V^{-1}\(\tau\)
\lab{calkdef}
\er  
By integrating \rf{eqforv2} from $\Sigma_{R}$  to $\Sigma$, one obtains $V$ evaluated on $\Sigma$, which is now an ordered volume integral, over the volume $\Omega$ inside $\Sigma$, and the ordering is determined by the scanning of $\Omega$ by closed surfaces as described above. But this result has of course to be the same as that obtained in \rf{firstvint}, i.e. by integrating \rf{eqforvb} when the surface  is closed, namely $\partial\Omega$.  Therefore, we obtain the generalized non-abelian Stokes theorem for a two-form connection $B_{\mu\nu}$, parallel transported by a one-form connection $A_{\mu}$  
\be
V_R\, P_2 \, e^{\int_{\partial\Omega}d\tau d\sigma W^{-1}B_{\mu\nu}W \frac{dx^{\mu}}{d\sigma}\,\frac{d\,x^{\nu}}{d\,\tau}}= P_3\, e^{\int_{\Omega} d\zeta \,{\cal K}}\,V_{R}
\lab{stokes}
\ee
where $P_3$ means volume ordering according to the scanning described above, and $V_R$ is  the integration constant obtained when integrating \rf{eqforvb} and \rf{eqforv2}. It corresponds in fact to the value of $V$ at the reference point $x_R$.   Note that such theorem holds true on a space-time of any dimension, and since the calculations leading to it make no mention to a metric tensor, it is valid on flat or curved space-time. The only restrictions appear when the topology of the space-time is non-trivial (existence of handles or holes for instance).  

\section{The construction of the integral equation for the full Yang-Mills theory}
\label{sec:proofym}
\setcounter{equation}{0}

One notes that \rf{basic} can be obtained from \rf{stokes} by replacing  $B_{\mu\nu}$  by $ie\left[ \alpha\, F_{\mu\nu}+\beta\, {\widetilde F}_{\mu\nu}\right]$, and using  the Yang-Mills equations
\be
D_{\nu}F^{\nu\mu}= J^{\mu} \qquad\qquad \qquad D_{\nu}{\widetilde F}^{\nu\mu}=0
\lab{diffymeq}
\ee  
to replace $\(D_{\lambda}B_{\mu\nu}+D_{\mu}B_{\nu\lambda}+D_{\nu}B_{\lambda\mu}\)$ in \rf{calkdef} by $(-ie\beta {\widetilde J}_{\mu\nu\lambda})$, and so ${\cal K}$ introduced in \rf{calkdef} is now given by
${\cal K}=\int_0^{2\,\pi}d\tau\, V\, {\cal J}\,V^{-1}$, 
with ${\cal J}$ given in \rf{caljdef}.
 Therefore, \rf{basic} is a direct consequence of the Yang-Mills equations \rf{diffymeq}  and the Stokes theorem \rf{stokes}. Note that  $V_R$  introduced in \rf{stokes}, does not appear in \rf{basic} because it has to lie in the  centre $Z\(G\)$ of $G$ to keep the gauge covariance of \rf{basic}. Indeed, consider a gauge transformation
\be
A_{\mu}\rightarrow g\, A_{\mu}\, g^{-1}+\frac{i}{e}\,\partial_{\mu}g\, g^{-1}\; ; 
\qquad  \qquad
F_{\mu\nu}\rightarrow g\, F_{\mu\nu}\,g^{-1} \; ;
\qquad \qquad
J_{\mu}\rightarrow g\, J_{\mu}\, g^{-1}
\lab{gaugetransf}
\ee
From \rf{eqforw},   $W\rightarrow g_f\, W\, g_i^{-1}$, with $g_i$ and $g_f$ being the values of $g$ at the initial and final points respectively of the path  determining $W$. Consequently,  ${\cal J}$ defined in \rf{caljdef} transforms as ${\cal J}\rightarrow g_R\, {\cal J}\, g_R^{-1}$, 
with $g_R$ being the value of $g$ at $x_R$. One also has $T\(A,\tau\)\rightarrow g_R\, T\(A,\tau\)\, g_R^{-1}$, and so from  \rf{eqforv}  $V \rightarrow g_R\, V\, g_R^{-1}$.  Similarly, one sees that ${\cal K}\rightarrow g_R\,{\cal K}\,g_R^{-1}$, and so  \rf{eqforv2} also implies that $V$ transforms as $V \rightarrow g_R\, V\, g_R^{-1}$. Note however that if $V_1$ is a solution of \rf{eqforv} so is $V_2=k\,V$ with $k$ being a constant element of  $G$. Similarly, if $V_3$ satisfies \rf{eqforv2} so does $V_4=V\, h$, with $h\in G$ being constant.  Under a gauge transformation  $V_1\rightarrow g_R\,V_1\,g_R^{-1}$, and  $V_2\rightarrow g_R\,V_2\,g_R^{-1}=g_R\,k\,V_1\,g_R^{-1}$. But $k$ is any chosen constant group element and it should not depend upon the gauge field, and so it  should not change under gauge transformations. In fact, the arbitrariness associated to $k$ corresponds to the choice of integration constants in \rf{eqforv} and \rf{eqforv2}. From this point of view we should have $V_2\rightarrow k\, g_R\, V_1\,g_R^{-1}$.  The only way to establish  the compatibility is to have $k\, g_R= g_R\,k$, i.e. $k$ should be an element of the centre $Z\(G\)$ of $G$. A similar analysis applies to $V_3$ and $V_4$. Therefore, the transformation law $V \rightarrow g_R\, V\, g_R^{-1}$, and so the gauge covariance of \rf{basic}, is only valid when the integration constants in \rf{eqforv} and \rf{eqforv2} are taken in $Z\(G\)$. In such case, $V_R$ cancels out of \rf{stokes} and that is why it does not appear in \rf{basic}. Consequently \rf{basic} transforms covariantly  under the gauge transformations \rf{gaugetransf}.

The integral equation \rf{basic} implies the local Yang-Mills equations. In order to see that, consider the case where $\Omega$ is an  infinitesimal volume of rectangular shape with lengths $dx^{\mu}$, $dx^{\nu}$ and $dx^{\lambda}$ along three chosen Cartesian axis labelled by $\mu$, $\nu$ and $\lambda$.  We choose the reference point $x_R$ to be at a vertex of $\Omega$. By considering only the lowest order contributions, in the lengths of $\Omega$, to the integrals in \rf{basic}, one observes that the surface and volume ordering become irrelevant. We have to pay attention only to the orientation of the derivatives of the coordinates w.r.t. the parameters $\sigma$, $\tau$ and $\zeta$, determined by the scanning of $\Omega$ described above. In addition, the contribution of a given face of $\Omega$ for the l.h.s. of \rf{basic} can be obtained by evaluating the integrand on any given point of the face since the differences will be of higher order.  Consider the two faces parallel to the plane $x^{\mu}x^{\nu}$. The contribution to the l.h.s. of \rf{basic} of the face at $x_R$ is   given by $-ie(\alpha F_{\mu\nu}+\beta{\widetilde F}_{\mu\nu})_{x_R}dx^{\mu}dx^{\nu}$, with the minus sign due to the orientation of the derivatives, and the contribution of the face at $x_R+dx^{\lambda}$ is 
$ie(W^{-1}(\alpha F_{\mu\nu}+\beta{\widetilde F}_{\mu\nu})W)_{(x_R+dx^{\lambda})}dx^{\mu}dx^{\nu}$, with $W_{(x_R+dx^{\lambda})}\sim \one-ieA_{\lambda}\(x_R\)dx^{\lambda}$. By Taylor expanding the second term, the joint contribution is 
$ieD_{\lambda}(\alpha F_{\mu\nu}+\beta{\widetilde F}_{\mu\nu})_{x_R}dx^{\mu}dx^{\nu}dx^{\lambda}$, with no sums in the Lorentz indices. The contributions of the other two pairs of faces are similar, and the l.h.s. of \rf{basic} to lowest order is $\one +ie(D_{\lambda}[\alpha F_{\mu\nu}+\beta{\widetilde F}_{\mu\nu}]+\mbox{cyclic perm.})_{x_R}dx^{\mu}dx^{\nu}dx^{\lambda}$.  When evaluating the r.h.s. of \rf{basic} we can take the integrand at any point of $\Omega$ since the differences are of higher order. In addition, the commutator term in ${\cal J}$ given in \rf{caljdef} is of higher order w.r.t. the first term involving the current. Therefore, the r.h.s. of \rf{basic} to lowest order is $\one+ie\beta{\widetilde J}_{\mu\nu\lambda}dx^{\mu}dx^{\nu}dx^{\lambda}$. Equating the coefficients of $\alpha$ and $\beta$ one gets  the pair of the (Hodge dual) Yang-Mills equations \rf{diffymeq}. 

\section{Path independency on loop space and the\\ conserved charges}
\label{sec:charges}
\setcounter{equation}{0}

Let us discuss some consequences of \rf{basic}. In order to write it for a given volume $\Omega$,  we had to choose a reference point $x_R$ on its border, and define a scanning of $\Omega$ with surfaces and loops. If one changes the reference point and the scanning, both sides of \rf{basic} will change. However, the generalized non-abelian Stokes theorem \rf{stokes} guarantees that the changes are such that both sides are still equal to each other. Therefore, one can say that \rf{basic} transforms ``covariantly'' under the change of scanning and reference point.  In fact to be precise, the equation \rf{basic} is formulated not on $\Omega$ but on the generalized loop space 
\be
L\Omega= \left\{ \gamma: S^2 \rightarrow \Omega\, |\, {\mbox{\rm north pole}} \rightarrow x_R\in \partial\Omega\right\}
\lab{lomegadef}
\ee
 The image of a given $\gamma$ is a closed surface $\Sigma$ in $\Omega$ containing $x_R$. A scanning of $\Omega$ is a collection of surfaces $\Sigma$, parametrized by $\tau$, such that $\tau=0$ corresponds to the infinitesimal surface around $x_R$ and $\tau =2\pi$ to $\partial\Omega$. Such collection of surfaces is a path in $L\Omega$ and each one corresponds to $\Omega$ itself. In order to perform each mapping $\gamma$ we scan the corresponding surface $\Sigma$ with closed loops starting and ending at $x_R$, and each loop is parametrized by $\sigma$, in the same way as we did in the arguments leading to \rf{stokes}. Therefore, the change of the scanning of $\Omega$ corresponds to a change of path in $L\Omega$. In this sense, the r.h.s. of \rf{basic} is a path dependent quantity in $L\Omega$ and its l.h.s. is evaluated at the end of the path. Of course, we do not want  physical quantities to depend upon the choice of paths in $L\Omega$, neither on the reference point. Note that if we take, in the four dimensional space-time $M$, a closed tridimensional volume $\Omega_c$, then the integral Yang-Mills equation \rf{basic} implies that 
\be
P_3e^{\oint_{\Omega_c} d\zeta d\tau  V{\cal J}V^{-1}}=\one
\lab{niceclosed}
\ee
since the border $\partial\Omega_c$ vanishes, and the ordered integral of the l.h.s. of \rf{basic} becomes trivial. On the loop space $L\Omega_c$, $\Omega_c$ corresponds to a closed path starting and ending at $x_R$. Consider now a point $\gamma$ on that closed path, corresponding to a closed surface $\Sigma$, in such a way that $\Omega_1$ corresponds to the first part of the path and $\Omega_2$  to the second, i.e. $\Omega_c= \Omega_1+ \Omega_2$, and $\Sigma$ is the common border of $\Omega_1$ and $\Omega_2$. By the ordering of the integration determined by  \rf{eqforv2} one observes that the relation \rf{niceclosed} can be split as $P_3e^{\int_{\Omega_2} d\zeta d\tau  V{\cal J}V^{-1}}P_3e^{\int_{\Omega_1} d\zeta d\tau  V{\cal J}V^{-1}}=\one$. However, by reverting the sense of integration along the path,  one gets the inverse operator when integrating \rf{eqforv2}. Therefore, $\Omega_1$ and $\Omega_2^{-1}$ are two different paths (volumes) joining the same points, namely  the infinitesimal surface around $x_R$ and the surface $\Sigma$, which correspond to their border. One then concludes that the operator $P_3e^{\int_{\Omega} d\zeta d\tau  V{\cal J}V^{-1}}$ is independent of the path, and so of the scanning of $\Omega$, as long as the end points, i.e. $x_R$ and the border $\partial\Omega$, are kept fixed.  

\subsection{The conserved charges for the full Yang-Mills theory}

The path independency of that operator can be used to construct conserved charges using the ideas of \cite{afs1,afs2}. First of all, let us assume that the space-time is of the form ${\cal S}\times \IR$, with $\IR$ being time and ${\cal S}$ the spatial sub-manifold which we assume simply connected and without border. An example is when ${\cal S}$ is the three dimensional sphere $S^3$.  It follows from \rf{niceclosed} that $Q_{{\cal S}}\equiv P_3e^{\oint_{{\cal S}} d\zeta d\tau  V{\cal J}V^{-1}}=\one$. That means that $Q_{{\cal S}}$ is not only  conserved in time,  but also that there can be no net charge  in ${\cal S}$. In fact, there is the possibility of getting charge quantization conditions in such case, if for some reason at the quantum level $\alpha$ and $\beta$ are not  free parameters. Indeed, take for instance Maxwell theory \cite{alvarez-olive}, where $G=U(1)$, and  so the commutators in \rf{caljdef} drop,  the surface and volume ordering are irrelevant, and  $Q_{{\cal S}}$ is unity if  
\be
\int_{{\cal S}}d\zeta d\tau d\sigma{\widetilde J}_{\mu\nu\lambda}
\frac{dx^{\mu}}{d\sigma}\frac{dx^{\nu}}{d\tau}
\frac{dx^{\lambda}}{d\zeta} = \frac{2\pi n}{e\beta}
\lab{chargequantization}
\ee
with $n$  integer.   At the classical level, where $\beta$ is a free parameter, the only acceptable solution to  \rf{chargequantization} is $n=0$, and so there should be no net charge is a space-time of the form ${\cal S}\times \IR$, with ${\cal S}$ being closed, i.e. with no border. 

\begin{figure}[ht]
\centering
\subfigure[]{
   \includegraphics[width=0.2\textwidth]{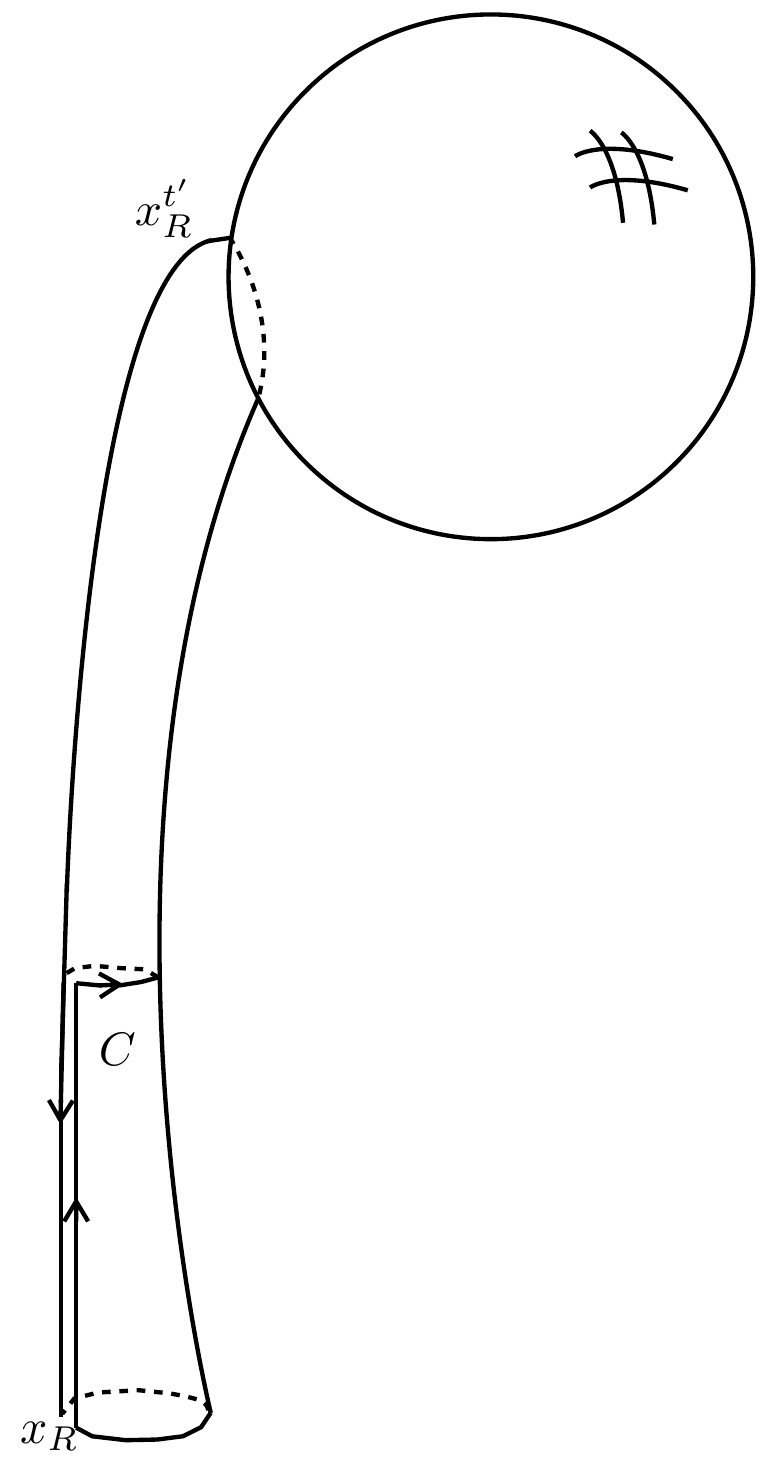}
   \label{closed_surface_a}
} \hspace{1cm}
 \subfigure[]{
   \includegraphics[width=0.2\textwidth]{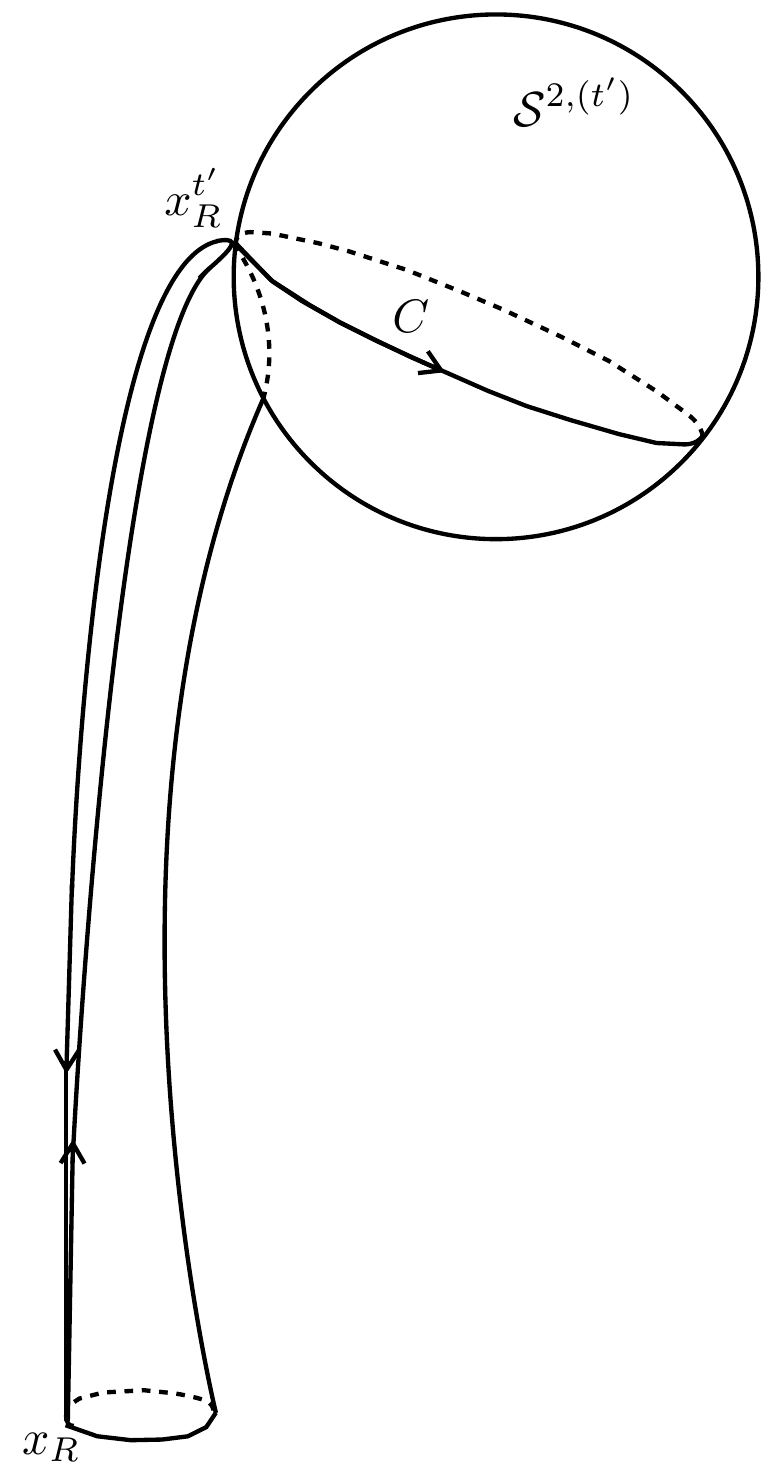}
   \label{closed_surface_b}
 } \hspace{1cm}
 \subfigure[]{
   \includegraphics[width=0.2\textwidth]{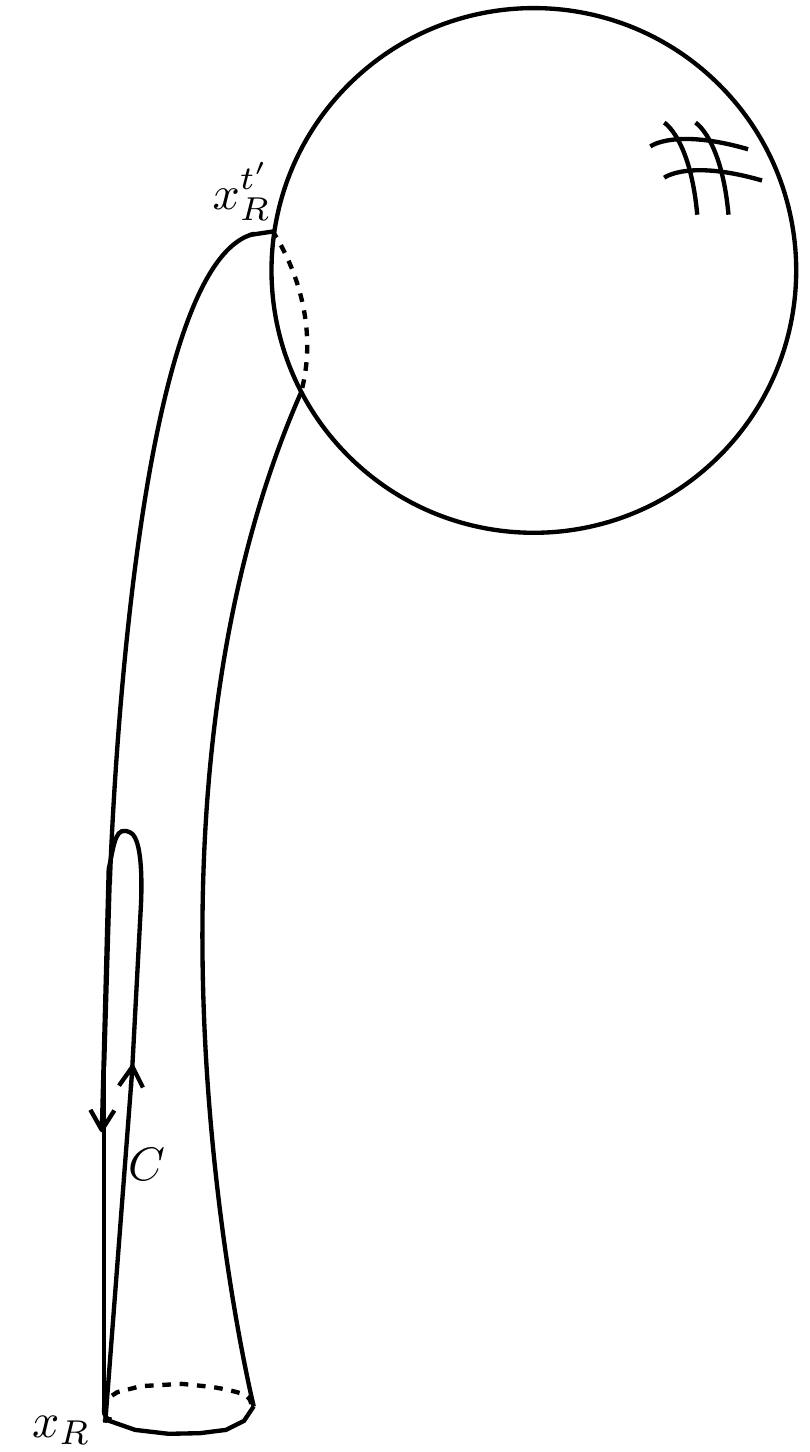}
   \label{closed_surface_c}
 }
\label{myfigure}
\caption{We  scan a hyper-cylinder $S^2\times I$ with surfaces of type shown above. Such surfaces are then scanned with loops as follows: as we go up the neck we scan it with loops as shown in figure (a), then the sphere $S^{2,(t^{\prime})}$ is scanned with loops as shown in figure (b), finally as we go down the neck we scan it with loops as shown in figure (c). In all cases the loops start and end at $x_R$. }
\end{figure}

Let us now assume that the space-time is not bounded, but still simply connected, like $\IR^4$. We shall consider two paths (volumes) joining the same two points, namely the infinitesimal surface around  $x_R$, which we take to be at the time $x^0=0$, and the two-sphere at spatial infinity $S^{2,(t)}_{\infty}$, at $x^0=t$. The first path is made of two parts. The first part corresponding to the whole space at  $x^0=0$, i.e. the volume $\Omega_{\infty}^{(0)}$ inside $S^{2,(0)}_{\infty}$, the two-sphere at spatial infinity  at $x^0=0$. The second part is a hyper-cylinder $S^2_{\infty}\times I$, where $I$ is the time interval between $x^0=0$ and $x^0=t$, and $S^2_{\infty}$ is a two-sphere at spatial infinity at the times on that interval. The second path is also made of two parts.  The first one corresponds to the infinitesimal hyper-cylinder $S^2_0\times I$, where $S^2_0$ is the infinitesimal two-sphere around $x_R$ and $I$ as before.  The second part  corresponds to $\Omega_{\infty}^{(t)}$, the whole space at time $x^0=t$, i.e. the volume inside $S^{2,(t)}_{\infty}$.  From the path independency following from \rf{niceclosed} one has that the integration of \rf{eqforv2}  along those two paths should give the same result, i.e.  
$V(S^2_{\infty}\times I)\,V(\Omega_{\infty}^{(0)})=V(\Omega_{\infty}^{(t)})V(S^2_0\times I)$, 
where we have used the notation $V\(\Omega\)\equiv P_3e^{\int_{\Omega} d\zeta d\tau  V{\cal J}V^{-1}}$, and where all integrations start at the reference point $x_R$ taken to be at $x^0=0$, and at the border $S^{2,(0)}_{\infty}$ of $\Omega_{\infty}^{(0)}$. In fact, one obtains $V\(\Omega\)$ by integrating \rf{eqforv2}, and so one has to calculate  ${\cal K}=\int_0^{2\,\pi}d\tau\, V\, {\cal J}\,V^{-1}$, on the surfaces scanning the volume $\Omega$. We shall scan a hyper-cylinder $S^2\times I$ with surfaces, based at $x_R$, of the form given in Figure 1, with $t^{\prime}$ denoting a time in the interval $I$. Each one of such surfaces are scanned with loops, labelled by $\tau$, in the following way. For $0\leq \tau\leq \frac{2\pi}{3}$, we scan the infinitesimal cylinder as shown in figure (1.a), then for $\frac{2\pi}{3}\leq \tau\leq \frac{4\pi}{3}$ we scan the sphere $S^2$ as shown in figure (1.b), and finally for $\frac{4\pi}{3}\leq \tau\leq 2\pi$ we go back to  $x_R$ with loops as shown in figure (1.c). The quantity ${\cal K}$ can then be split into the contributions coming from each one of those surfaces as ${\cal K}={\cal K}_a+{\cal K}_b+{\cal K}_c$. In the case of the infinitesimal hyper-cylinder $S^2_0\times I$, the sphere has infinitesimal radius and so it does not really contribute to ${\cal K}_b$. We shall assume the currents and field strength vanish at spatial infinity as
$$
J_{\mu} \sim \frac{1}{R^{2+\delta}}
\qquad \qquad{\rm and} \qquad  \qquad
F_{\mu\nu}\sim \frac{1}{R^{\frac{3}{2}+\delta^{\prime}}}
$$
with $\delta,\delta^{\prime}>0$, for $R\rightarrow \infty$. Therefore the quantity ${\cal J}$, given in \rf{caljdef}, vanishes when calculated on loops at spatial infinity. Consequently, in the case of the hyper-cylinder $S^2_{\infty}\times I$, the contribution to ${\cal K}_b$ coming from the sphere with infinite radius vanishes, and we have that ${\cal K}$ calculated on the surfaces scanning $S^2_{\infty}\times I$ and $S^2_0\times I$ is the same, and so $V\(S^2_{\infty}\times I\)=V\(S^2_0\times I\)$. In fact there is more to it, since when we contract the radius of the cylinders in Figure 1 to zero the loops in figures (1.a) and (1.c) become the same. Therefore, the quantities ${\cal J}$ calculated on them are the same except for a minus sign coming from the derivatives $\frac{dx^{\mu}}{d\tau}$, since the loops in figure (1.a) get longer  with the increase of $\tau$, and in figure (1.c) the opposite occurs. In addition, the quantity $V$ inside the the expression  ${\cal K}=\int_0^{2\,\pi}d\tau\, V\, {\cal J}\,V^{-1}$ is insensitive to that sign since it is obtained by integrating \rf{eqforv} starting at $x_R$ in both cases. Therefore, it turns out that ${\cal K}_a+{\cal K}_c=0$. The loops scanning the sphere in figure (1.b) have legs linking the reference point $x_R$, at $x^0=0$, to the same space point but at $x^0=t^{\prime}$, i.e. $x_R^{t^{\prime}}$. Therefore, when integrating \rf{eqforv} one gets $V_{x_R}= W(x_R^{t^{\prime}},x_R)^{-1}V_{x_R^{t^{\prime}}}W(x_R^{t^{\prime}},x_R)$, where $W(x_R^{t^{\prime}},x_R)$ is obtained by integrating \rf{eqforw} along the leg linking $x_R$ to $x_R^{t^{\prime}}$, and where we have used the notation $V_x$, meaning $V$ obtained from \rf{eqforv} with reference point $x$. Using the same arguments and notation one obtains from \rf{caljdef} that, on the loops of figure (1.b), ${\cal J}_{x_R}= W(x_R^{t^{\prime}},x_R)^{-1}{\cal J}_{x_R^{t^{\prime}}}W(x_R^{t^{\prime}},x_R)$, and so  ${\cal K}_{b,x_R}= W(x_R^{t^{\prime}},x_R)^{-1}{\cal K}_{b,x_R^{t^{\prime}}}W(x_R^{t^{\prime}},x_R)$. The quantity $V(\Omega_{\infty}^{(t)})$ is obtained by integrating \rf{eqforv2} and by scanning the volume $\Omega_{\infty}^{(t)}$ with surfaces of the type shown in figure (1.b), and where the radius of $S^2$ varies from zero to infinity keeping the point $x_R^t$ fixed. Therefore, from the above arguments one gets that 
\be
V_{x_R}(\Omega_{\infty}^{(t)})=W(x_R^{t},x_R)^{-1}V_{x_R^{t}}(\Omega_{\infty}^{(t)})W(x_R^{t},x_R)
\lab{moverefpoint}
\ee
 One then concludes that such operator has an iso-spectral time evolution 
\be
V_{x_R^{t}}(\Omega_{\infty}^{(t)})=U(t)V_{x_R}(\Omega_{\infty}^{(0)})U(t)^{-1}
\qquad \quad{\rm with} \qquad \quad U(t)=W(x_R^{t},x_R)V\(S^2_{0}\times I\)
\lab{conservedoperatorfullym}
\ee
Therefore, its eigenvalues, or equivalently ${\rm Tr}(V_{x_R^{t}}(\Omega_{\infty}^{(t)}))^N$, are constant in time. Note that from the Yang-Mills equations \rf{basic} one has that such operator can be written either as a volume or surface ordered integrals, and so we have proved \rf{charge}. 

Note that if $V_{x_R^{t}}(\Omega_{\infty}^{(t)})$ has an iso-spectral evolution so does $g_cV_{x_R^{t}}(\Omega_{\infty}^{(t)})$, with $g_c \in Z(G)$, the center of the gauge group. That fact has to do with the freedom we have to choose the integration constants of \rf{eqforv} and \rf{eqforv2} to lie in $Z(G)$, without spoiling the gauge covariance of \rf{basic} (see discussion in the proof of \rf{basic} above).

{\bf Properties of the charges.}   First of all we point out that the conserved charges are gauge invariant. Indeed,  using the same arguments given below \rf{gaugetransf} for the proof of \rf{basic},   one has that under the gauge transformations \rf{gaugetransf} the operator $V_{x_R^{t}}$ transforms as
$$
  V_{x_R^{t}}(\Omega_{\infty}^{(t)})\rightarrow g_RV_{x_R^{t}}(\Omega_{\infty}^{(t)})g_R^{-1}
  $$ 
  with $g_R$ being  the group element, performing the gauge transformation, at $x_R^{t}$. Therefore its eigenvalues, which are the charges, are gauge invariant. Note that the operator $V_{x_R^{t}}$ is the same as that given in \rf{charge}, since the volume $\Omega_{\infty}^{(t)}$ corresponds to the spatial sub-manifold $S$ at time equals $t$.
  
Note that when one changes the reference point   from $x_R^t$ to ${\widetilde x}_R^t$, the operator $V_{x_R^{t}}(\Omega_{\infty}^{(t)})$ changes under conjugation by $W({\widetilde x}_R^t,x_R^t)$, i.e. the holonomy of the gauge field $A_{\mu}$ through a path joining those two points. Therefore, similarly to \rf{moverefpoint} one has
$$
V_{x_R^{t}}(\Omega_{\infty}^{(t)}) \rightarrow W({\widetilde x}_R^t,x_R^t)^{-1}\, V_{x_R^{t}}(\Omega_{\infty}^{(t)})\, W({\widetilde x}_R^t,x_R^t)
$$
So the conserved quantities, being the eigenvalues of $V_{x_R^{t}}(\Omega_{\infty}^{(t)}) $, are also independent of the base points.   Note in addition that the reference points $x_R^t$ and ${\widetilde x}_R^t$ are on the border of the volume $\Omega_{\infty}^{(t)}$ and so they lie at spatial infinity.   Our boundary conditions imply that the field strength goes to zero at infinity and so the gauge potential is asymptotically flat, and consequently $W({\widetilde x}_R^t,x_R^t)$ is independent of the choice of path joining the two reference points.  

We have seen below \rf{lomegadef} that the volume $\Omega$ can be seen as a path in  the loop space $L\Omega$. In fact, there is an infinite number of paths in $L\Omega$ representing the same physical volume $\Omega$, due to the infinite ways of scanning $\Omega$ with closed surfaces based at $x_R$. We have shown that, as a consequence of \rf{niceclosed}, the operator $P_3e^{\int_{\Omega} d\zeta d\tau  V{\cal J}V^{-1}}$ is independent of the path on loop space,  as long as the end points, i.e. $x_R$ and the border $\partial\Omega$, are kept fixed.  When we say that we have to keep the end points $x_R$ and $\partial\Omega$ fixed, we mean that not only the physical point $x_R$ and surface $\partial\Omega$ are kept fixed but also its scanning with loops. We do not have to worry about the scanning of $x_R$ because that is trivial. A re-parameterization of the volume $\Omega$ corresponds to a change of the path in loop space representing $\Omega$. Therefore, the operator $V_{x_R^{t}}(\Omega_{\infty}^{(t)})$, or equivalently the r.h.s. of \rf{charge} is independent of the re-parameterization of the volume $\Omega_{\infty}^{(t)}$. Consequently, the conserved charges, which are the eigenvalues of that operator, are invariant under re-parameterization (scanning) of the volume $\Omega_{\infty}^{(t)}$ that keep fixed its end points, i.e. keep fixed the physical point $x_R$ and surface $\partial \Omega_{\infty}^{(t)}$, as well as its scanning with loops. 

We now have to analyze  how the charges transform when we fix the physical point $x_R$ and surface $\partial \Omega_{\infty}^{(t)}$, but change the scanning of $\partial \Omega_{\infty}^{(t)}$ with loops. Again we do not have to worry about the scanning of the infinitesimal surface around $x_R$ because that is trivial.  The volume $\Omega_{\infty}^{(t)}$ corresponds to the spatial sub-manifold $S$ introduced in \rf{charge}, at time equals $t$. Therefore, we have to analyze how the surface ordered integral over $\partial S$ given in \rf{charge}, transforms under the change of the scanning of $\partial S$ with loops.  Remember however that such integral is obtained by integrating \rf{eqforv} over $\partial S$. In \rf{deltav} we have shown how such kind of integral changes when the surface of integration is changed. The calculation in \rf{deltav} is valid not only for a change of the physical surface but also for a change of the scanning of it with loops. In the latter case the variation $\delta x^{\mu}\(\sigma\)$ of the loop is parallel to the surface, i.e. there are no variations $\delta x^{\mu}\(\sigma\)$ perpendicular to the surface. Since $\partial S$ is a two-dimensional surface all $3$-forms on it vanish. Therefore, the first term of  \rf{deltav} must vanish trivially when we restrict $\delta x^{\lambda}$ to be parallel to the surface, since $\frac{d\,x^{\mu}}{d\,\sigma}$ and $\frac{d\,x^{\nu}}{d\,\tau}$ are, by definition, parallel to the surface. The argument we are using here is the same as that in the proof of Theorem 2.1 of \cite{afs2} for $r$-flat connections in loop space.  Replacing $B_{\mu\nu}$  by $ie\left[ \alpha\, F_{\mu\nu}+\beta\, {\widetilde F}_{\mu\nu}\right]$ we get that \rf{eqforvb} becomes \rf{eqforv} and \rf{tdef}. Therefore, the condition for the surface ordered integral over $\partial S$ in \rf{charge}, to be invariant under changes of the scanning of $\partial S$  with loops is 
\br
 &&\int_0^{2\,\pi}d\sigma\,\int_0^{\sigma}d\sigma^{\prime}
\sbr{\(\alpha-1\)\, F_{\kappa\rho}^W\(\sigma^{\prime}\)+ \beta\, {\widetilde F}_{\kappa\rho}^W\(\sigma^{\prime}\)}
{\alpha\, F_{\mu\nu}^W\(\sigma\)+\beta\, {\widetilde F}_{\mu\nu}^W\(\sigma\)}
\nonumber\\
&& \times \,\frac{dx^{\kappa}}{d\sigma^{\prime}}\frac{dx^{\mu}}{d\sigma}
\(\frac{d\,x^{\rho}\(\sigma^{\prime}\)}{d\,\tau}\delta x^{\nu}\(\sigma\)
-\delta x^{\rho}\(\sigma^{\prime}\)\,\frac{d\,x^{\nu}\(\sigma\)}{d\,\tau}\)=0
\lab{newcondition}
\er  
where we have used the notation $X^W\equiv W^{-1}\, X\, W$.  Such double integral in $\sigma$ and $\sigma^{\prime}$ is performed over a given loop, based at $x_R$,  scanning the surface $\partial S$. Note however that such surface is the border of the spatial sub-manifold $S$, and so it lies at spatial infinity. Therefore, the field tensor and its Hodge dual, appearing in the integrand, are to be evaluated at spatial infinity. 

There are at least two sufficient conditions for \rf{newcondition} to hold true. The first one is that the field tensor and its dual should fall at spatial infinity faster than $1/R^2$,  where $R$ is the radius (or a measure of size) of the surface $\partial S$, which should be taken to infinity, i.e. $R\rightarrow \infty$. That is so, because the integrand in \rf{newcondition} is quartic in variations of the Cartesian coordinates $x^{\mu}$ and quadratic in the field tensor and its dual. As we will see in section \ref{sec:instantons} that is exactly what happens in the case of instantons. So, the conserved (in the Euclidean time) charges constructed as the eigenvalues   of the operator \rf{charge} are invariant under re-parameterization of volumes and surfaces for the instantons solutions. 

The second sufficient condition is that the commutator in the integrand of \rf{newcondition} must vanish, and so the field tensor and its dual conjugated by the holonomy $W$ has to lie in an abelian subalgebra. As we discuss in sections \ref{sec:monopole} and \ref{sec:merons} that is exactly what happens in the cases of monopoles, dyons and merons.  For those solutions the field tensor at spatial infinity has the form $F_{\mu\nu} \sim \frac{1}{r^2}\, G\({\hat r}\)$, where $r$ is the radial distance, and $G\({\hat r}\)$ is a Lie algebra element depending on the unit vector ${\hat r}$, and being covariantly constant, i.e. $D_{\mu} G\({\hat r}\)=0$. That fact implies that $W^{-1}\, G\({\hat r}\)\, W$ is constant everywhere, and so all components of the field tensor and its dual (conjugated by $W$) lies in the direction of that element of the Lie algebra. Therefore, the commutator in  the integrand of \rf{newcondition} vanishes. Consequently, the conserved  charges constructed as the eigenvalues   of the operator \rf{charge} are invariant under re-parameterization of volumes and surfaces for the monopole, dyon and meron solutions.

\subsubsection{Comparing with the textbook conserved charges}

The usual conserved charges for the non-abelian gauge theories discussed in textbooks are essentially those proposed by  Yang and Mills in their original paper \cite{ym}, and they are constructed as follows. Using Yang-Mills equations \rf{diffymeq} one introduces  the currents
\be
j^{\mu}\equiv \partial_{\nu}F^{\nu\mu}= J^{\mu}-ie\sbr{A_{\nu}}{F^{\nu\mu}} \qquad\qquad\qquad
{\widetilde j}^{\mu}\equiv \partial_{\nu}{\widetilde F}^{\nu\mu}= -ie\sbr{A_{\nu}}{{\widetilde F}^{\nu\mu}}
\ee
which are conserved due to the antisymmetry of the field tensor, i.e. $\partial_{\mu}j^{\mu}=0$ and   $\partial_{\mu}{\widetilde j}^{\mu}=0$. Under appropriate boundary conditions, the corresponding conserved charges are given by 
\be
Q_{YM}=\int d^3x\,\partial_{i}F^{i0}= \int_{S^2_{\infty}} d{\vec \Sigma}\cdot {\vec E} 
\qquad\qquad\qquad\quad
{\widetilde Q}_{YM}=\int d^3x\,\partial_{i}{\widetilde F}^{i0}= -\int_{S^2_{\infty}} d{\vec \Sigma}\cdot {\vec B}
\lab{wrongcharge}
\ee
where $S^2_{\infty}$ is a two-sphere at spatial infinity, and $E_i\equiv F_{0i}$, $B_i \equiv -\frac{1}{2}\,\varepsilon_{ijk}\,F_{jk}$, are the non-abelian electric and magnetic fields respectively. Under the gauge transformations \rf{gaugetransf} one has
\be
Q_{YM}\rightarrow \int_{S^2_{\infty}} d{\vec \Sigma}\cdot g\, {\vec E}\, g^{-1}
\qquad\qquad\qquad\quad
{\widetilde Q}_{YM}\rightarrow -\int_{S^2_{\infty}} d{\vec \Sigma}\cdot g\,{\vec B}\, g^{-1}
\ee
If one restricts oneself to gauge transformations where the group element $g$ goes to a constant at spatial infinity then the  charges transform covariantly, i.e. $Q_{YM}\rightarrow g_{\infty}\,Q_{YM}\,g_{\infty}^{-1}$, and ${\widetilde Q}_{YM}\rightarrow g_{\infty}\,{\widetilde Q}_{YM}\,g_{\infty}^{-1}$, with $g_{\infty}$ being the constant group element on $S^2_{\infty}$. Therefore, the eigenvalues of $Q_{YM}$ and ${\widetilde Q}_{YM}$ are invariant under those restricted gauge transformation. 

The conserved charges we construct in this paper, namely the eigenvalues of the operator \rf{charge} differ in many aspects from the charges \rf{wrongcharge}. First, we show that only the eigenvalues of the operator \rf{charge} are conserved in time. The full operator has an isospectral time evolution. Second, those eigenvalues are invariant under general gauge transformations, and not only under restricted transformations where the group element goes to a constant at infinity. Third, the charges obtained from  \rf{charge} are different from those obtained from \rf{wrongcharge}. Indeed, as we show in the example of the monopole in section \ref{sec:monopole}, all the charges coming from \rf{wrongcharge} vanish, and those obtained from \rf{charge} give the magnetic charge of the monopole, as well as its quantization. There are two aspects to stress here. First our calculations work equally well for the Wu-Yang monopole as well as for the 'tHooft-Polyakov monopole, and the conservation of the charge is dynamical and not necessarily topological. Second, the magnetic charge which is conserved is associated to the abelian subgroup only (in the case the gauge group is $SO(3)$, to the $U(1)$ subgroup) and not to the full group as is the case of the charge coming from  \rf{wrongcharge}. That is also true for the Wu-Yang monopole where the gauge symmetry is not spontaneously broken.  For those reasons, the construction of conserved charges for non-abelian gauge theories that we propose in this paper constitute a great advance with respect to what is usually found in the literature.

\subsection{The conserved charges for the self-dual sector}  
 
 The integral equation \rf{integralselfdualym}  leads to some interesting consequences which we now discuss. Consider the case where the surface $\Sigma$ is closed, i.e. it has no border. Then, $\partial \Sigma$ disappears  and the l.h.s. of \rf{integralselfdualym} is trivial and so 
\be
P_2\,e^{ie\int_{\partial\Omega}d\sigma d\tau\,W^{-1}\left[\alpha\,F_{\mu\nu}+\kappa\,\(1-\alpha\)\,{\widetilde F}_{\mu\nu}\right]W\frac{dx^{\mu}}{d\sigma}\frac{dx^{\nu}}{d\tau}}
=\one
\lab{niceselfdual}
\ee
where we have denoted $\Sigma =  \partial\Omega$, i.e. the border of the volume $\Omega$ contained inside $\Sigma$.  Now, if one takes $\beta=\kappa\(1-\alpha\)$ in \rf{basic}, one observes that the second term in the expression \rf{caljdef} for ${\cal J}$ vanishes, when the self dual equation \rf{integralselfdualym}, or equivalently \rf{selfdualym}, are imposed. Therefore, \rf{niceselfdual} and \rf{basic} imply that 
\be
P_3\,e^{ie\kappa\(1-\alpha\) \int_{\Omega} d\zeta d\tau  d\sigma V
   {\widetilde J}_{\mu\nu\lambda}^W
\frac{dx^{\mu}}{d\sigma}\frac{dx^{\nu}}{d\tau}
\frac{dx^{\lambda}}{d\zeta} 
V^{-1}}=\one
\ee
Since that has to be valid on any volume $\Omega$, one concludes that the current $J_{\mu}$ should vanish. That is an expected result, and indeed the imposition of the Yang-Mills equations \rf{diffymeq} and the self duality equations \rf{selfdualym} imply the vanishing of $J_{\mu}$. In addition, the first order differential equations \rf{selfdualym} imply the second order equations \rf{diffymeq}  when the current vanishes, since the second equation in \rf{diffymeq} is just an identity, the so-called Bianchi identity, i.e.
\be
D_{\lambda}F_{\mu\nu}+D_{\mu}F_{\nu\lambda}+D_{\nu}F_{\lambda\mu}=0
\lab{bianchi}
\ee 
In order to understand that the integral self dual Yang-Mills equation \rf{integralselfdualym} implies the the full integral Yang-Mills equation \rf{basic} we have to construct the integral version of the Bianchi identity. One can obtain that by taking the generalized non-abelian Stokes theorem \rf{stokes}, which is an identity, and choose $B_{\mu\nu}=i\,e\,\lambda\, F_{\mu\nu}$, with $\lambda$ a free parameter, and use \rf{bianchi} to get
\be
P_2 e^{ie\,\lambda \,\int_{\partial\Omega}d\tau d\sigma F_{\mu\nu}^W\frac{dx^{\mu}}{d\sigma}\frac{dx^{\nu}}{d\tau}}=  P_3e^{\int_{\Omega} d\zeta d\tau  V{\cal C}V^{-1}}
\lab{nontrivial}
\ee
with 
\br
{\cal C}&\equiv&
e^2\,\lambda\,\(\lambda-1\)\,
 \int_0^{2\pi}d\sigma \int_0^{\sigma}d\sigma^{\prime}
\sbr{ F_{\kappa\rho}^W\(\sigma^{\prime}\)}
{ F_{\mu\nu}^W\(\sigma\)} 
\nonumber\\
&& \times
\, \frac{d\,x^{\kappa}}{d\,\sigma^{\prime}}\frac{d\,x^{\mu}}{d\,\sigma}
\(\frac{d\,x^{\rho}\(\sigma^{\prime}\)}{d\,\tau}\frac{d\,x^{\nu}\(\sigma\)}{d\,\zeta}
-\frac{d\,x^{\rho}\(\sigma^{\prime}\)}{d\,\zeta}\frac{d\,x^{\nu}\(\sigma\)}{d\,\tau}\)
\lab{calcdef}
\er  
The relation \rf{nontrivial} is highly non-trivial for $\lambda \neq 0 \; {\rm or}\; 1$. Indeed, for $\lambda=1$ it leads to what one would naively expect as the integral version of the Bianchi identity, i.e.  $P_2 e^{ie\int_{\partial\Omega}d\tau d\sigma F_{\mu\nu}^W\frac{dx^{\mu}}{d\sigma}\frac{dx^{\nu}}{d\tau}}=\one$. The relation \rf{nontrivial} carries important information about the flux of a non-abelian field strength $F_{\mu\nu}$ through a closed surface when that is rescaled by a factor $\lambda$, and it certainly  deserves further investigation. Now, by imposing the self duality equation \rf{integralselfdualym}, or equivalently \rf{selfdualym}, one observes that \rf{basic} becomes \rf{nontrivial} with $\lambda=\alpha+\kappa \beta$. So, the self duality condition does turn the full integral Yang-Mills equation \rf{basic} into an identity, namely \rf{nontrivial}, in a manner similar that \rf{selfdualym} does to the full differential Yang-Mills equation \rf{diffymeq}.

The other consequence of the relation \rf{niceselfdual} is that it leads to conservation laws, in a manner similar to that which \rf{niceclosed} does in the case of the full Yang-Mills equations. In order to do that let us consider the loop space associated to a surface $\Sigma$
\be
L\Sigma=\{ \gamma : S^1 \rightarrow \Sigma \mid \mbox{\rm north pole} \rightarrow x_R \in \partial \Sigma\}
\ee
A scanning of $\Sigma$ with loops, in the way described below  \rf{integralselfdualym}, corresponds to a path in $L\Sigma$. In fact, there is an infinity of paths in $L\Sigma$ corresponding to the same physical surface $\Sigma$. The closed surface $\partial \Omega$ is a closed path in $L\partial \Omega$, and the reference point $x_R$ is now any chosen point on $\partial \Omega$ since it has no border. Let us now take a point in that closed path corresponding to a  loop in $\partial \Omega$. It then splits the path into two parts, or equivalently  $\partial \Omega$ into two surfaces with a common border, i.e. $\partial \Omega=\Sigma_1+\Sigma_2$. Consequently \rf{niceselfdual} can be written as 
\be
P_2\,e^{ie\int_{\Sigma_1}d\sigma d\tau\,W^{-1}\left[\alpha\,F_{\mu\nu}+\kappa\,\(1-\alpha\)\,{\widetilde F}_{\mu\nu}\right]W\frac{dx^{\mu}}{d\sigma}\frac{dx^{\nu}}{d\tau}} \; 
P_2\,e^{ie\int_{\Sigma_2}d\sigma d\tau\,W^{-1}\left[\alpha\,F_{\mu\nu}+\kappa\,\(1-\alpha\)\,{\widetilde F}_{\mu\nu}\right]W\frac{dx^{\mu}}{d\sigma}\frac{dx^{\nu}}{d\tau}}
=\one
\lab{niceselfdual2}
\ee
By reverting the sense of integration along the path one gets the inverse operator. But $\Sigma_1$ and $\Sigma_2^{-1}$ are two different paths joining the same two points in the loop space $L\partial \Omega$, namely the infinitesimal loop around $x_R$ and the common border of  $\Sigma_1$ and $\Sigma_2^{-1}$. Therefore, \rf{niceselfdual2} implies that the operator  
\be
V\(\Sigma\)\equiv 
P_2\,e^{ie\int_{\Sigma}d\sigma d\tau\,W^{-1}\left[\alpha\,F_{\mu\nu}+\kappa\,\(1-\alpha\)\,{\widetilde F}_{\mu\nu}\right]W\frac{dx^{\mu}}{d\sigma}\frac{dx^{\nu}}{d\tau}}
\lab{vsigma}
\ee
 is independent of the path, or equivalently of the surface $\Sigma$, as long as the end points (the border of $\Sigma$ and the reference point $x_R$ on it) are kept fixed. In addition, by fixing the surface $\Sigma$ and changing the path in the loop space $L\Sigma$ that corresponds to it, we see that $V\(\Sigma\)$ is independent of the choice of the scanning. Such path independency leads to conservation laws as we now explain. 

 First of all let us fix a plane in an Euclidean space-time $M$, and let  us denote by $x^{\mu}$ and $x^{\nu}$  ($\mu$ and $\nu$ fixed)  the Cartesian coordinates associated to the two orthogonal axis lying on that plane. That could be done in a space time of any metric, but we are interested in real solutions of the self dual Yang-Mills equations \rf{selfdualym}, and so we choose the Euclidean metric. We shall denote by $x^{\alpha}$ and $x^{\beta}$ ($\alpha$ and $\beta$ fixed) the two Cartesian coordinates corresponding to the directions orthogonal to the plane $x^{\mu}\,x^{\nu}$. In fact, we shall work with a given fixed axis parameterized by $t$ which is a linear combination of the $x^{\alpha}$ and $x^{\beta}$ axis, i.e. we write 
\be
x^{\alpha}= t\, \cos \phi \qquad\qquad \qquad x^{\beta}= t\, \sin \phi \qquad\qquad \qquad -\infty < t < \infty \qquad 0\leq \phi \leq \pi
\lab{taxisdef}
\ee
 \begin{figure}[t]
  \begin{center}
    \includegraphics[width=0.5\textwidth]{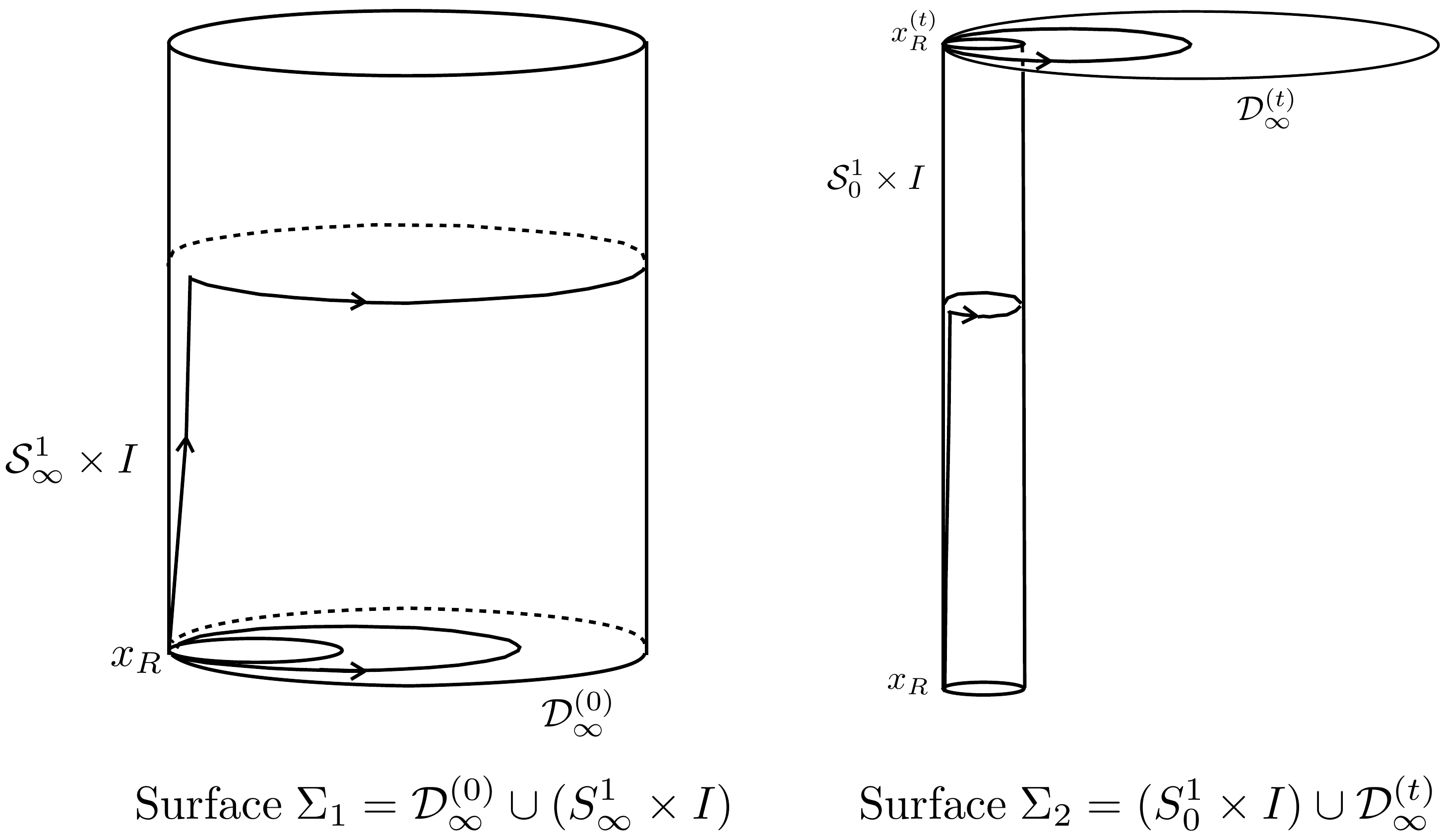}
  \end{center}
  \caption{The surfaces $\Sigma_1$ and $\Sigma_2$, with the same border 
  $S^{1,(t)}_{\infty}$, and reference point $x_R$, used in the construction of conserved charges.}
 \label{fig:surface_scan} 
\end{figure}
We shall choose two surfaces $\Sigma_1$ and $\Sigma_2$ with the same border as shown in Figure \ref{fig:surface_scan}. The surface $\Sigma_1$ is made of two parts. The first part is a disc ${\cal D}_{\infty}^{(0)}$ of infinite radius on the plane $x^{\mu}\,x^{\nu}$ at $t=0$, i.e. it is the whole plane $x^{\mu}\,x^{\nu}$ at $t=0$.  The second part is a cylinder $S_{\infty}^1\times I$, where $I$ is a segment of the $t$-axis going from $t=0$ to $t=t$, and $S_{\infty}^1$ is a circle of infinite radius parallel to the plane $x^{\mu}\,x^{\nu}$. We choose the reference point $x_R$ to lie on the border of ${\cal D}_{\infty}^{(0)}$, i.e. on the circle $S_{\infty}^1$ at $t=0$. The surface $\Sigma_2$ is also made of two parts. The first part is an infinitesimal cylinder $S_{0}^1\times I$ , with $I$ as before, and $S_{0}^1$ a circle of infinitesimal radius also parallel to the plane $x^{\mu}\,x^{\nu}$. We choose the infinitesimal circle $S_{0}^1$ such that $x_R$ lies on it at $t=0$. The second part of $\Sigma_2$ is a disc ${\cal D}_{\infty}^{(t)}$ of infinite radius, parallel to the plane $x^{\mu}\,x^{\nu}$, and at $t=t$. We scan the two surfaces with loops, as shown in Figure \ref{fig:surface_scan}, starting and ending at the reference point $x_R$ (for a similar discussion on how to do that see section 3.1 of \cite{second}). Therefore, the surfaces $\Sigma_1$ and $\Sigma_2$ are two different paths in the loop space $L\partial \Omega$, such that $\partial \Omega= \Sigma_1+\Sigma_2^{-1}$, with the same end points, namely the infinitesimal circle at $x_R$ and the circle $S_{\infty}^1$ at $t=t$. Since the operator \rf{vsigma} is independent of the surface, it follows that it is the same calculated on those two surfaces, i.e.
\be
V\(\Sigma_1\)= V\(\Sigma_2\)\qquad\quad \rightarrow \qquad\quad 
V\({\cal D}_{\infty}^{(0)}\)V\(S_{\infty}^1\times I\) = V\(S_{0}^1\times I\)V\({\cal D}_{\infty}^{(t)}\)
\ee
Note that in fact, $V\(S_{0}^1\times I\)=\one$ since $S_{0}^1$ is infinitesimal. Now if we impose the boundary conditions 
\be
F_{\rho\sigma}=\kappa\, {\widetilde F}_{\rho\sigma} \sim \frac{1}{r^{2+\delta}}\, T\({\hat r}\) \qquad \qquad \qquad{\rm for}\qquad \qquad r\rightarrow \infty
\lab{euclideanboundarycond}
\ee
where $\delta >0$, $r$ is the radial distance in the $x^{\mu}\,x^{\nu}$ plane, i.e. $r^2=\(x^{\mu}\)^2+\(x^{\nu}\)^2$ ($\mu$ and $\nu$ fixed),  and $T\({\hat r}\)$ is a Lie algebra element depending only on the radial direction, i.e. ${\hat r}=\frac{{\vec r}}{r}$. Those boundary conditions imply that the integrand in \rf{vsigma} vanishes on the cylinder $S_{\infty}^1\times I$, and so $V\(S_{\infty}^1\times I\) =\one$. Therefore, one gets that 
$V\({\cal D}_{\infty}^{(0)}\) = V\({\cal D}_{\infty}^{(t)}\)$. We can not say yet we have a conserved quantity in the parameter $t$, because both operators are calculated using the same reference point $x_R$ at $t=0$. Let now $x_R^{(t)}$ be a point with the same $x^{\mu}$ and $x^{\nu}$ coordinates but at $t=t$, i.e. lying at the border of ${\cal D}_{\infty}^{(t)}$ (see Figure \ref{fig:surface_scan}). If we now scan ${\cal D}_{\infty}^{(t)}$ with loops based at $x_R^{(t)}$ we get an operator $V_{x_R^{(t)}}\({\cal D}_{\infty}^{(t)}\)$, which is related to that, based at $x_R$, as
$V_{x_R}\({\cal D}_{\infty}^{(t)}\) = W^{-1}\(x_R^{(t)},x_R\)\,V_{x_R^{(t)}}\({\cal D}_{\infty}^{(t)}\)\,W\(x_R^{(t)},x_R\)$, 
where $W\(x_R^{(t)},x_R\)$ is the holonomy of the gauge potential $A_{\mu}$, obtained by integrating \rf{eqforw}, along the line joining $x_R$ to $x_R^{(t)}$, and where the subindices,  $x_R$ and $x_R^{(t)}$, indicate the reference point using in the calculation of the operator \rf{vsigma}. Therefore, one gets that
\be
V_{x_R^{(t)}}\({\cal D}_{\infty}^{(t)}\)  = W\(x_R^{(t)},x_R\)\,V_{x_R}\({\cal D}_{\infty}^{(0)}\)\,W^{-1}\(x_R^{(t)},x_R\)
\ee
From \rf{vsigma} and the integral self dual Yang-Mills equation \rf{integralselfdualym} we get that
\be
V_{x_R^{(t)}}\({\cal D}_{\infty}^{(t)}\)  =
P_2\,e^{ie\int_{{\cal D}_{\infty}^{(t)}}d\sigma d\tau\,W^{-1}\left[\alpha\,F_{\mu\nu}+\kappa\,\(1-\alpha\)\,{\widetilde F}_{\mu\nu}\right]W\frac{dx^{\mu}}{d\sigma}\frac{dx^{\nu}}{d\tau}}
= P_1\,e^{-ie\oint_{S_{\infty}^{1,(t)}}d\sigma\, A_{\mu}\frac{dx^{\mu}}{d\sigma}}
\lab{constantoperator}
\ee
where $S_{\infty}^{1,(t)}$ is the border of ${\cal D}_{\infty}^{(t)}$. 

The result we have obtained is that the operator \rf{constantoperator}  has an iso-spectral evolution in $t$. Then, its eigenvalues, or equivalently ${\rm Tr}\left[V_{x_R^{(t)}}\({\cal D}_{\infty}^{(t)}\)\right]^N$, are constant in $t$. But by rotating the axis $t$ (see \rf{taxisdef}), one gets that those eigenvalues are constant on the whole plane $x^{\alpha}\,x^{\beta}$. But since $V_{x_R^{(t)}}\({\cal D}_{\infty}^{(t)}\)$ is integrated over the whole plane $x^{\mu}\,x^{\nu}$, it turns out that the eigenvalues are in fact independent of all coordinates of the Euclidean space-time $M$. In addition, such construction is independent of the choice of the orientation of the plane $x^{\mu}\,x^{\nu}$. We stress that such conserved charges are gauge invariant, independent of the parametrization of the surfaces and also of the choice of the reference point $x_R$. The arguments for such facts are similar to those presented in the case of the conserved charges of the full Yang-Mills equations (see paragraph below \rf{conservedoperatorfullym}). We have therefore proved the relation \rf{chargeselfdual}.

\subsubsection{Interpreting the charges \rf{chargeselfdual}}
\label{sec:interpretation}

Note that the proof of \rf{chargeselfdual}, or equivalently \rf{constantoperator}, was based on the equation \rf{niceselfdual} following from the integral self-dual equation \rf{integralselfdualym}. However, if one takes \rf{niceselfdual} on shell, i.e. when \rf{selfdualym} holds true, then \rf{niceselfdual} becomes
\be
P_2\,e^{ie\int_{\partial\Omega}d\sigma d\tau\,W^{-1}\,F_{\mu\nu}\,W\frac{dx^{\mu}}{d\sigma}\frac{dx^{\nu}}{d\tau}}
=\one
\lab{niceselfdualidentity}
\ee
But that is just an identity following either from the usual non-abelian Stokes theorem \rf{usualstokes} by taking $C_{\mu}\equiv ie\, A_{\mu}$, and $\Sigma$ a closed surface, i.e. the border of a volume $\Sigma=\partial\Omega$, or then from the integral Bianchi identity \rf{nontrivial} with $\lambda=1$. Therefore, if the field tensor satisfies the boundary conditions (see \rf{euclideanboundarycond}) 
\be
F_{\rho\sigma} \sim \frac{1}{r^{2+\delta}}\, T\({\hat r}\) \qquad \qquad \qquad{\rm for}\qquad \qquad r\rightarrow \infty
\lab{euclideanboundarycondidentity}
\ee
with $\delta >0$, all the arguments leading to \rf{constantoperator} hold true, and we obtain an isospectral evolution for the operator
\be
V^{\prime}_{x_R^{(t)}}\({\cal D}_{\infty}^{(t)}\)  =
P_2\,e^{ie\int_{{\cal D}_{\infty}^{(t)}}d\sigma d\tau\,W^{-1}\,F_{\mu\nu}\,W\frac{dx^{\mu}}{d\sigma}\frac{dx^{\nu}}{d\tau}}
= P_1\,e^{-ie\oint_{S_{\infty}^{1,(t)}}d\sigma\, A_{\mu}\frac{dx^{\mu}}{d\sigma}}
\lab{constantoperatoridentity}
\ee
Therefore the eigenvalues of \rf{constantoperatoridentity} are constant in the time $t$ introduced in \rf{taxisdef}. Note that such result applies to any field configuration satisfying 
\rf{euclideanboundarycondidentity} and it does not have necessarily to be a self-dual solution of the Yang-Mills equations. In fact, it does not even  have to be a solution of the Yang-Mills theory since \rf {niceselfdualidentity} follows from identities. 

Note that if ${\cal D}_{\infty}^{(t)}$ is a spatial surface then the surface ordered integral in \rf{constantoperatoridentity}, namely $P_2\,e^{ie\int_{{\cal D}_{\infty}^{(t)}}d\sigma d\tau\,W^{-1}\,F_{\mu\nu}\,W\frac{dx^{\mu}}{d\sigma}\frac{dx^{\nu}}{d\tau}}$, corresponds to the flux of the non-abelian magnetic field ($B_i \equiv -\frac{1}{2}\,\varepsilon_{ijk}\,F_{jk}$) through that surface.  On the other hand, if ${\cal D}_{\infty}^{(t)}$ has a time component then that integral corresponds to the flux of the non-abelian electric field ($E_i\equiv F_{0i}$) through such spatial-temporal surface. Note that the conservation of those fluxes can be intuitively understood by the fact that the border of ${\cal D}_{\infty}^{(t)}$ is the circle $S_{\infty}^{1,(t)}$ of infinite radius. Therefore, if the field configuration is localized in a region at a finite distance to the plane containing $S_{\infty}^{1,(t)}$ the solid angle defined by that circle is $2\,\pi$ spheroradians. If that field configuration evolves in the time $t$ changing its distance to that plane by a finite amount, the solid angle will remain the same, and so should the flux of the magnetic or electric fields. Of course, that is an intuitive view, and so not precise, of the conservation of the charge, but we will show that it stands reasonable in the examples we discuss in section \ref{sec:examples}.

\section{Examples}
\label{sec:examples}
\setcounter{equation}{0}

We now evaluate the conserved charges obtained from \rf{charge} and \rf{chargeselfdual} for well known solutions like monopoles, dyons, instantons and merons. For simplicity we restrict ourselves to the case where the gauge group is $SU(2)$, since it contains all the physically relevant aspects of the construction. 

\subsection{Monopoles and dyons}
\label{sec:monopole}

In order to evaluate the operator \rf{charge} let us first work with its form as a surface ordered integral of the field tensor and its dual, and then consider the volume ordered integral form of it. Therefore we need the field tensor at spatial infinity only. The 'tHooft-Polyakov \cite{thooft} and Wu-Yang \cite{wuyang} monopoles for a gauge group $SU(2)$ have the same behavior at infinity. Indeed, the gauge field and field tensor at infinity are given by 
\br
 A_i &=&   -\frac{1}{e}\,\varepsilon_{ijk}\, \frac{{\hat r}_j}{r}\, T_k=\frac{1}{2}\,\frac{i}{e}\,\partial_i  g\,g^{-1} \; ;  \qquad\qquad \qquad A_0=0 
\nonumber\\
F_{ij}&=&\frac{1}{e}\,\varepsilon_{ijk}\, \frac{{\hat r}_k}{r^2}\; {\hat r}\cdot T
\; ;  \qquad\qquad \qquad \qquad\qquad \quad F_{0i} =0
\lab{monopole}
\er
where ${\hat r}=\frac{{\vec r}}{r}$, is unit vector in the radial direction, $T_i$ are the generators of the $SU(2)$ Lie algebra satisfying
\be
\sbr{T_i}{T_j}=i\,\varepsilon_{ijk}\,T_k \qquad\qquad \qquad\qquad i,j,k=1,2,3
\lab{su2}
\ee 
and $g$ is the group element $g=\exp\(i\,\pi\,{\hat r}\cdot T\)$. In the case of the Wu-Yang monopole the formulas \rf{monopole} correspond to the exact solution, and not only to its behavior at infinity. In the case of the 'tHooft-Polyakov monopole, on the other hand, \rf{monopole} is true only in the limit $r\rightarrow \infty$, and we do not show the behavior of the Higgs field since it is not relevant in the evaluation of the charges as we show below.

In order to calculate \rf{charge} we have to scan the two-sphere at spatial infinity  with loops starting and ending at a chosen reference point $x_R$. The quantity $W$ is obtained by integrating \rf{eqforw} from $x_R$ to a given point on the loop.  An important fact in such calculation is that the quantity ${\hat r}\cdot T$ is covariantly constant, i.e.
\be
D_i \,{\hat r}\cdot T= \partial_i\,{\hat r}\cdot T+i\,\,e\, \sbr{A_i}{{\hat r}\cdot T}=0
\lab{rtcov}
\ee
Therefore, using \rf{eqforw} one gets that
\be
\frac{d\;}{d\sigma} \(W^{-1}\,{\hat r}\cdot T\,W\)=0
\lab{rtconstant}
\ee
So, $W^{-1}\,{\hat r}\cdot T\,W$ is constant along any loop, and consequently constant everywhere. If we denote by $T_R$ the value of ${\hat r}\cdot T$ at the reference point $x_R$, one gets from \rf{monopole} that 
\be
W^{-1}\,F_{ij}\,W = \frac{1}{e}\,\varepsilon_{ijk}\, \frac{{\hat r}_k}{r^2}\; T_R
\lab{nicefijtr}
\ee
and so it belongs to the abelian subalgebra $U(1)$ generated by $T_R$. Therefore, the surface ordering becomes irrelevant and the operator \rf{charge} becomes (since ${\widetilde F}_{ij}=0$)
\br
Q_S&=& e^{i\,e\,\alpha\,\int_{S^2_{\infty}}d\sigma\,d\tau\,W^{-1}\,F_{ij}\,W\,\frac{dx^i}{d\sigma}\,\frac{dx^j}{d\tau} }=e^{-i\,e\,\alpha\,\int_{S^2_{\infty}}d{\vec\Sigma}\cdot{\vec B}^R }
\nonumber\\
&=& e^{i\,\alpha\,T_R\,\int_{S^2_{\infty}}d\sigma\,d\tau\,\varepsilon_{ijk}\, \frac{{\hat r}_k}{r^2}
 \,\frac{dx^i}{d\sigma}\,\frac{dx^j}{d\tau} }=e^{i\,4\,\pi\,\alpha\,T_R}
 \lab{qsmonopole}
\er
where we have introduced the abelian magnetic field $B^R_i\equiv -\frac{1}{2}\,\varepsilon_{ijk}\, W^{-1}\,F_{jk}\,W= -\frac{1}{e}\, \frac{{\hat r}_i}{r^2}\, T_R$, and have denoted $d\Sigma_i=\varepsilon_{ijk}\, \frac{dx^j}{d\sigma}\,\frac{dx^k}{d\tau}\,d\sigma\,d\tau$. Using Gauss law we define the magnetic charge as
\be
\int_{S^2_{\infty}}d{\vec\Sigma}\cdot{\vec B}^R =  G_R \qquad\qquad \mbox{\rm and so}\qquad \qquad 
G_R= -\frac{4\,\pi}{e}\, T_R
\ee
According to our construction (see \rf{charge}) the eigenvalues of $Q_S$ are constant in time which, in view of \rf{qsmonopole}, is equivalent to say that the eigenvalues of $G_R$ are constants. At the end of section \ref{sec:statements}, where we discuss the nature of the eigenvalues of the charges, we have shown that our construction does not fix the vector space (representation) where such eigenvalues should be evaluated. If we choose to calculate them on a finite dimensional representation of the gauge group $SU(2)$ (or $SO(3)$), then the eigenvalues of $T_R$ are integers or half-integers. Therefore it follows that the magnetic charges, on those representations, must be quantized as 
\be
\mbox{\rm eigenvalues of $G_R$}= \frac{2\,\pi\,n}{e} \qquad\qquad \qquad n=0,\pm 1,\pm 2\ldots
\lab{quantizemagnetic}
\ee
Let us now look at the evaluation of the magnetic charges as volume ordered integrals. From  \rf{charge}, \rf{caljdef} and \rf{qsmonopole} one gets that 
\be
e^{-i\,e\,\alpha\,G_R}= P_3e^{\int_{{\rm space}} d\zeta d\tau  V{\cal J}_{{\rm monopole}}V^{-1}}
\lab{monopolevolumeformula}
\ee
with
\br
{\cal J}_{{\rm monopole}}&\equiv&
e^2\,\alpha\,\(\alpha-1\)\,
 \int_0^{2\pi}d\sigma \int_0^{\sigma}d\sigma^{\prime}
\sbr{ F_{ i j}^W\(\sigma^{\prime}\)}
{ F_{ k l}^W\(\sigma\)} 
\nonumber\\
&& \times
\, \frac{d\,x^{ i}}{d\,\sigma^{\prime}}\frac{d\,x^{ k}}{d\,\sigma}
\(\frac{d\,x^{ j}\(\sigma^{\prime}\)}{d\,\tau}\frac{d\,x^{ l}\(\sigma\)}{d\,\zeta}
-\frac{d\,x^{ j}\(\sigma^{\prime}\)}{d\,\zeta}\frac{d\,x^{ l}\(\sigma\)}{d\,\tau}\)
\lab{calcdejmonopole}
\er 
where we have used the fact that  ${\widetilde F}_{ij}=0$, and  ${\widetilde J}_{123}=J_0=0$, since in the Wu-Yang case there is no current, and in the 'tHooft-Polyakov case we have a static solution with $A_0=0$, and so the time component of the Higgs field current vanishes.  Note that \rf{monopolevolumeformula} and \rf{calcdejmonopole} could also have been obtained from the integral Bianchi identity \rf{nontrivial}.  In the case of the 'tHooft-Polyakov monopole it follows that \rf{nicefijtr} is not true inside the monopole core and we have a quite non-trivial expression for the magnetic charge as a volume integral. We do not evaluate it in this paper and so we do not have anything to add to the result \rf{quantizemagnetic}. Note however, that even though the r.h.s. of \rf{monopolevolumeformula} is integrated over the entire space, the Higgs field does not contribute for such formula of the magnetic charge. 

In the case of the Wu-Yang monopole, however, it is not that difficult to evaluate \rf{calcdejmonopole} after performing a regularization of the Wilson line, passing through the singularity of the gauge potential \rf{monopole}. That calculation is given in the appendix \ref{sec:wilson} and the result is that ${\cal J}_{{\rm monopole}}$ vanishes in all loops, and so (see \rf{unityvolumeintwuyang})
\be
 e^{-i\,e\,\alpha\,G_R}=P_3e^{\int_{{\rm space}} d\zeta d\tau  V{\cal J}_{{\rm monopole}}V^{-1}}=\one 
\ee
Such result implies that the magnetic charge for the Wu-Yang monopole is quantized as
\be
\mbox{\rm eigenvalues of $G_R$}= \frac{2\,\pi\,n}{e\, \alpha} \qquad\qquad \qquad n=0,\pm 1,\pm 2\ldots
\lab{quantizemagneticwuyang}
\ee
If the parameter $\alpha$ is indeed arbitrary, and there is no physical condition to fix it, then the only acceptable value for the integer $n$ is $n=0$, and so the magnetic charge of the Wu-Yang monopole should vanish.  Perhaps we have to go to the quantum theory to settle that issue. It might happen that quantum conditions restrict the allowed values of $\alpha$. That is one of the important points of our construction to be further investigated.

Let us now consider the case  of dyon solutions.  For the Wu-Yang and the 'tHooft-Polyakov case, as calculated by Julia and Zee \cite{julia}, the space components of the gauge potential and field tensor, namely $A_i$ and $F_{ij}$, $i,j=1,2,3$, are the same as those in \rf{monopole}, and the time components, at spatial infinity, are replaced by
\be
A_0= \frac{M}{e}\, {\hat r}\cdot T+ \frac{\gamma}{e}\,\frac{{\hat r}\cdot T}{r}+ O(\frac{1}{r^2})\; ;
\qquad\qquad 
F_{0i}= \frac{\gamma}{e}\,\frac{{\hat r}_i}{r^2}\,{\hat r}\cdot T + O(\frac{1}{r^3})\; ;
\qquad\qquad r\rightarrow \infty
\lab{dyon}
\ee
with $M$ and $\gamma$ being parameters of the solution. In the case of the Wu-Yang dyon, i.e. when there is no Higgs field and no symmetry breaking, the formulas \rf{dyon}, as well as \rf{monopole}, are true everywhere and not only at spatial infinity. In other words, there are no terms of order $r^{-2}$ and $r^{-3}$ in $A_0$ and $F_{0i}$ respectively. 
Using \rf{rtconstant} we have, in anology to \rf{nicefijtr}, that 
\be
W^{-1}\,{\widetilde F}_{ij}\, W\rightarrow -\frac{\gamma}{e}\,\varepsilon_{ijk}\,\frac{{\hat r}_k}{r^2}\,T_R \qquad\qquad \qquad r\rightarrow \infty
\ee
So, $W^{-1}\,{\widetilde F}_{ij}\, W$ also belongs to the abelian subalgebra $U(1)$ generated by $T_R$, and it is in fact proportional to $W^{-1}\,F_{ij}\, W$. Therefore, the surface ordering is not relevant in the evaluation of the operator \rf{charge}, and we get in the dyon case that
\br
Q_S=e^{-i\,e\,\left[\alpha\,\int_{S^2_{\infty}}d{\vec\Sigma}\cdot{\vec B}^R +\beta\, 
\int_{S^2_{\infty}}d{\vec\Sigma}\cdot{\vec E}^R\right]}= 
e^{-i\,e\,\left[\alpha\,G_R +\beta\, K_R\right]}=
e^{i\,4\,\pi\,\left[\alpha -\beta\, \gamma\right]\,T_R}
\lab{dyonchargecons}
\er
where we have introduced the abelian electric field $E_i^R=W^{-1}\,F_{0i}\,W=\frac{\gamma}{e}\,\frac{{\hat r}_i}{r^2}\, T_R$,  ${\vec B}^R$ and $G_R$ are the same as before, and using Gauss law we have defined the electric charge as 
\be
\int_{S^2_{\infty}}d{\vec\Sigma}\cdot{\vec E}^R= K_R \qquad\qquad \mbox{\rm and so} \qquad\qquad K_R= \frac{4\,\pi\,\gamma}{e}\, T_R
\lab{krdef}
\ee
According to \rf{charge} the eigenvalues of $Q_S$ are constant in time, and so we conclude from \rf{dyonchargecons} that the eigenvalues of $\(\alpha\,G_R +\beta\, K_R\)$ are constants. But if we assume that the parameters $\alpha$ and $\beta$ are arbitrary it follows that the eigenvalues of $G_R$ and $K_R$ are indenpendently constant in time. We have seen that, by evaluating  the eigenvalues  of  $T_R$  on finite dimensional representations of the gauge group $SU(2)$, where they are integers or half-integers, the eigenvalues of the magnetic charge  $G_R$ are quantized as in \rf{quantizemagnetic}. Under the same assumptions it follows from \rf{krdef} that 
\be
\mbox{\rm eigenvalues of $K_R$}= \frac{2\,\pi\,\gamma\, n}{e}\qquad\qquad \qquad n=0,\pm1, \pm 2, \dots
\lab{quantizeelectric}
\ee
Again from \rf{charge} we can express the charges in terms of volume ordered integrals, and from \rf{charge}, \rf{caljdef} and \rf{dyonchargecons} we get 
\be
e^{-i\,e\,\left[\alpha\,G_R+\beta\, K_R\right]}= P_3e^{\int_{{\rm space}} d\zeta d\tau  V{\cal J}_{{\rm dyon}}V^{-1}}
\lab{dyonvolumeformula}
\ee
with
\br
{\cal J}_{{\rm dyon}}&\equiv&
\int_0^{2\pi}d\sigma\left\{ ie\beta {\widetilde J}_{ijk}^W
\frac{dx^{i}}{d\sigma}\frac{dx^{j}}{d\tau}
\frac{dx^{k}}{d\zeta} \right. \nonumber\\
&+& \left. e^2\int_0^{\sigma}d\sigma^{\prime}
\sbr{\(\(\alpha-1\) F_{ i j}^W+\beta {\widetilde F}_{ i j}^W\)\(\sigma^{\prime}\)}
{\(\alpha F_{ k l}^W+\beta {\widetilde F}_{ k l}^W\)\(\sigma\)} \right.
\nonumber\\
&&\left. \times
\, \frac{d\,x^{ i}}{d\,\sigma^{\prime}}\frac{d\,x^{ k}}{d\,\sigma}
\(\frac{d\,x^{ j}\(\sigma^{\prime}\)}{d\,\tau}\frac{d\,x^{ l}\(\sigma\)}{d\,\zeta}
-\frac{d\,x^{ j}\(\sigma^{\prime}\)}{d\,\zeta}\frac{d\,x^{ l}\(\sigma\)}{d\,\tau}\)\right\}
\lab{caljdyondef}
\er  
In the case of the Wu-Yang dyon we have ${\widetilde J}_{123}=J_0=0$, since there are no sources. However,  for the Julia-Zee dyon we have that the Higgs field contributes to the current $J_{\mu}$. One can extract the magnetic and electric charges $G_R$ and $K_R$ from \rf{dyonvolumeformula}, by setting $\beta=0$ and $\alpha=0$ respectively. Then, the Higgs field contributes  to the electric charge only.

In the case of the Wu-Yang dyon it is possible to evaluate \rf{caljdyondef} after a regularization of the Wilson line operator passing through the singularity of the gauge potential \rf{monopole}. That calculation is shown in the appendix \ref{sec:wilson} and it was found that ${\cal J}_{{\rm dyon}}$ vanishes in all loops for the Wu-Yang dyon (see \rf{unityvolumeintwuyang}). Therefore
\be
e^{-i\,e\,\left[\alpha\,G_R+\beta\, K_R\right]}= P_3e^{\int_{{\rm space}} d\zeta d\tau  V{\cal J}_{{\rm dyon}}V^{-1}}=\one
\lab{dyonvolumeformulawuyang}
\ee
which implies that
\be
\mbox{\rm eigenvalues of $\left[\alpha\,G_R+\beta\, K_R\right]$}= \frac{2\,\pi\, n}{e}\qquad\qquad \qquad n=0,\pm1, \pm 2, \dots
\lab{grkrquantization}
\ee
Again, if the parameters $\alpha$ and $\beta$ are indeed arbitrary, then it follows from \rf{grkrquantization} that by taking $\beta=0$, the eigenvalues of $G_R$ should obey  \rf{quantizemagneticwuyang}. On the other hand by taking $\alpha=0$, one concludes that \rf{grkrquantization} implies that
\be
\mbox{\rm eigenvalues of $K_R$}= \frac{2\,\pi\,n}{e\, \beta} \qquad\qquad \qquad n=0,\pm 1,\pm 2\ldots
\lab{quantizeelectricwuyang}
\ee
Now, if \rf{quantizemagneticwuyang} and \rf{quantizeelectricwuyang} should hold true for arbitrary values of $\alpha$ and $\beta$ respectively, then the only acceptable value of the integer $n$ in both equations is $n=0$, and consequently the electric and magnetic charges of the Wu-Yang dyon should vanish. As discussed below \rf{quantizemagneticwuyang}, we have perhaps to consider  of the quantum theory to settle that issue, since there could be quantum conditions restricting the values of $\alpha$ and $\beta$.

It is worth evaluating the conserved charges associated to the operator \rf{constantoperatoridentity} in the case of the monopole and dyon solutions. For simplicity we shall take the circle of infinite radius $S^{1,(t)}_{\infty}$ to lie on the plane $x^1\,x^2$, for some constant values of $x^3$ and $x^0$. The calculation for any other plane is similar and leads, as we shall see, to similar results. We use polar coordinates on the plane, with the polar angle being the parameter $\sigma$ parameterizing the circle, i.e. $x^1=\rho \, \cos \sigma$, $x^2=\rho \, \sin \sigma$, and  $r^2=\rho^2+\(x^3\)^2$, with $x^3$ constant, and $\rho\rightarrow \infty$. Therefore, for both the monopole and dyon solutions, we get from \rf{monopole} that on the circle of infinite radius we have $A_{\mu}\,\frac{dx^{\mu}}{d\sigma}=\frac{1}{e}\,T_3$, since on that circle $\rho\sim r\rightarrow \infty$ and the unit vector ${\hat r}$, on that limit,  has components only on the plane $x^1\,x^2$. Therefore, from \rf{constantoperatoridentity}, we have 
\be
V^{\prime}_{x_R^{(t)}}\({\cal D}_{\infty}^{(t)}\)  = P_1\,e^{-ie\oint_{S_{\infty}^{1,(t)}}d\sigma\, A_{\mu}\frac{dx^{\mu}}{d\sigma}}=e^{-i\,2\,\pi\,T_3}
\lab{halflux}
\ee
As shown in the arguments leading to \rf{constantoperatoridentity} the eigenvalues of such operator are conserved in the time $t$ which in this case can be any linear combination of $x^0$ and $x^3$. But that is equivalent to say that  the eigenvalues of $T_3$ are conserved in $t$. Since those are integers or half integers in a finite dimensional representation of $SU(2)$, it follows that the operator $V^{\prime}_{x_R^{(t)}}\({\cal D}_{\infty}^{(t)}\) $ is either $\one$ or $-\one$, i.e. an element of the center of $SU(2)$. As pointed out below \rf{constantoperatoridentity} such conserved charge can be interpreted as the non-abelian magnetic flux through the surface ${\cal D}_{\infty}^{(t)}$, which border is  $S_{\infty}^{1,(t)}$. Indeed, we see that the argument of the exponential in \rf{halflux} is half of the argument of the exponential in  \rf{qsmonopole}, if one takes $\alpha=1$, and considers that $T_R$ and $T_3$ have the same norm and so the same eigenvalues. Remember $T_R$ is the value of ${\hat r}\cdot T$ at the reference point $x_R$ and so ${\rm Tr}\(T_R\)^2={\rm Tr}\(T_i\,T_j\)\,{\hat r}_i\,{\hat r}_j=\lambda$, where ${\rm Tr}\(T_i\,T_j\)=\lambda\, \delta_{ij}$, and $\lambda$  depends upon the representation used. Since the argument of the exponential in  \rf{qsmonopole} corresponds to the total flux of the magnetic field through $S^2_{\infty}$, 
we see that it is the double of the flux through ${\cal D}_{\infty}^{(t)}$.   Due to the spherical symmetry of the solution that is compatible with  interpretation, given below \rf{constantoperatoridentity}, since $S^2_{\infty}$ corresponds to a solid angle of $4\,\pi$ spheroradians as seen from the center of the solution and ${\cal D}_{\infty}^{(t)}$ corresponds to only $2\pi$ spherodradians. 

There are several comments that are important to make regarding the construction of charges for monopoles and dyons. First of all the charges we constructed are different from those given by \rf{wrongcharge}. Indeed, from \rf{monopole} and \rf{dyon} we have that the magnetic and electric fields at spatial infinity for the Wu-Yang and 'tHooft-Polyakov cases are given by
\be
B_i\rightarrow -\frac{1}{e}\, \frac{{\hat r}_i}{r^2}\, {\hat r}\cdot T \; ;\qquad\qquad\qquad
 E_i\rightarrow \frac{\gamma}{e}\, \frac{{\hat r}_i}{r^2}\, {\hat r}\cdot T\; ; \qquad\qquad \qquad r \rightarrow \infty
\ee
So, they do not lie on an abelian $U(1)$ subalgebra like $B_i^R$ and $E_i^R$ given above, and when integrated on the two-sphere at infinity lead to the vanishing of the charges   \rf{wrongcharge}, i.e.
\be
Q_{YM}^{\rm monopole/dyon}={\widetilde Q}_{YM}^{\rm monopole/dyon}=0
\ee
Note that even though the evaluation of the charges \rf{qsmonopole} and \rf{dyonchargecons}
rely on the choice of a reference point $x_R$, which leads to a particular generator $T_R$, the charges do not depend upon that reference point. Indeed, if one changes the reference point from $x_R$ to ${\widetilde x}_R$, then the operator $Q_S$ changes as (see discussion below \rf{conservedoperatorfullym})  
\be
Q_S\rightarrow W({\widetilde x}_R,x_R)^{-1} \, Q_S\, W({\widetilde x}_R,x_R)
\ee
where $W({\widetilde x}_R,x_R)$ is the holonomy from the old reference point $x_R$ to the new one ${\widetilde x}_R$. Therefore the charges, which are the eigenvalues of $Q_S$, do not change. 

Note that since the charges are the eigenvalues of the operator \rf{charge}, the number of charges is equal to the rank of the gauge group $G$. However, since the field tensor and its Hodge dual come multiplied by the arbitrary  parameters $\alpha$ and $\beta$ respectively, the number of charges is in fact twice the rank of the gauge group. So, we have rank of $G$ magnetic charges and rank of $G$ electric charges. In this sense the number of charges does not pay attention to the pattern of symmetry breaking. Indeed, our calculations have shown that the electric and magnetic charges are the same for the Wu-Yang case, which is a solution of the pure Yang-Mills theory,  and for the 'tHooft-Polyakov case which has a Higgs field breaking the gauge symmetry from $SO(3)$ down to $SO(2)$. In fact, as we have shown above the Higgs field does not play any role in the evaluation of the charges.  In addition the conservation of the charges is dynamical, i.e. it follows directly from the integral form of the equations of motion \rf{basic}. That contrasts to the conservation of the magnetic charge of the 'tHooft-Polyakov monopole which follows from topological (homotopy) considerations related to the mapping of the Higgs field from the spatial infinity to the Higgs vacua. Another point relates to the quantization of the magnetic charge, which in the case of   'tHooft-Polyakov monopole comes from the topology again, i.e. the charge is determined by second homotopy group of the Higgs vacua. In our case, the quantization of the charges comes from the integral equations of motion themselves (more precisely the integral Bianchi identity \rf{nontrivial}), without any reference to the Higgs field since it works equally well for the Wu-Yang and  'tHooft-Polyakov monopoles. It is worth pointing out that the magnetic charges of monopoles of  'tHooft-Polyakov type have already been expressed in the literature, as surface ordered integral using the ordinary non-abelian Stokes theorem. See for instance section 5 of Goddard and Olive's review paper \cite{olive}. However, that construction is totally based on the  properties of the Higgs vacua, since the fact that the Higgs field must be covariantly constant at spatial infinity leads to an equation for it similar to \rf{eqforw} for the Wilson line $W$. In addition  the argument for the conservation of the magnetic charge is particular to that type of solution since it is based on topology considerations of the solution. The generalized non-abelian Stokes theorem \rf{stokes}, and consequently the integral Yang-Mills equations \rf{basic} were  not known by that time. We believe that the role played by the integral equation \rf{basic} in monopole and dyon solutions deserves further study specially in the quantum theory. It might connect to the so-called abelian projection and arguments for confinement.

 \subsection{Euclidean solutions}
 \label{sec:instantons}
 
 Note that the proof that the eigenvalues of the operator \rf{charge} are constant in the time $x^0$ did not really depend on the particular properties of the metric on the  Minkowski space-time. In fact, the metric tensor was only necessary  to introduce the Hodge duals of the field tensor and of the matter current. Therefore, one could evaluate those eigenvalues for Euclidean solutions, like instantons and merons, and obtain charges conserved in the Euclidean time $x^4$. We will see that those charges are trivial in the case of instantons but not in the case of merons, where they relate to magnetic type charges.   We shall also evaluate the charges associated to the operator \rf{constantoperatoridentity}, or equivalently \rf{chargeselfdual}, for the case of instantons and merons. 
 
  \subsubsection{Instantons}  
  
The instantons are euclidean self-dual solutions where the gauge potentials  become of the pure gauge form at infinity, i.e. $A_{\mu}\rightarrow \frac{1}{e}\,\partial_{\mu}g\,g^{-1}$, for $s\rightarrow \infty$, with $s^2= x_1^2+x_2^2+x_3^2+x_4^2$, where $x^{\mu}$, $\mu=1,2,3,4$, being the Cartesian coordinates in the Euclidean space-time.   That fact  simplifies many of the calculations, and makes trivial the eigenvalues of the operators  \rf{charge} and \rf{chargeselfdual}. However, as pointed out in the section \ref{sec:interpretation} the physical charges may be related to the eigenvalues of the Lie algebra elements associated by exponentiation to the group elements corresponding to the operators \rf{charge} and \rf{chargeselfdual}. We shall illustrate that with the cases of the one and two-instanton solutions.
 
Let us consider the case where the gauge group is $SU(2)$ and take the one-instanton solution \cite{manton} given by
\be
A_{\mu}=-\frac{2}{e}\,\sigma_{\mu\nu}\, \frac{x^{\nu}-a^{\nu}}{\(x^{\rho}-a^{\rho}\)^2+\lambda^2}\; ;
\qquad\qquad\quad
F_{\mu\nu}=\kappa\,{\widetilde F}_{\mu\nu}=\frac{\sigma_{\mu\nu}}{e}\,\frac{4\,\lambda^2}{\left[\(x^{\rho}-a^{\rho}\)^2+\lambda^2\right]^2}
\lab{one-instanton}
\ee 
with $\kappa=\pm 1$, and where $a^{\mu}$, $\mu=1,2,3,4$,  and $\lambda$ are parameters of the solution, and 
\be
\sigma_{i4}= -\kappa\, T_i\; ; \qquad\qquad \quad\sigma_{ij}=\varepsilon_{ijk}\, T_k\; ;
\qquad\qquad\quad \sbr{T_i}{T_j}=i\, \varepsilon_{ijk}\, T_k
\lab{sigmamunudef}
\ee
with $i,j,k=1,2,3$,  $T_i$ being the generators of the $SU(2)$ Lie algebra, and the quantities $\sigma_{\mu\nu}$ satisfy $\frac{1}{2}\varepsilon_{\mu\nu\alpha\beta}\,\sigma_{\alpha\beta}=\kappa\,\sigma_{\mu\nu} $. 

If one considers a $2$-sphere $S^2_{\infty}$ of infinite radius surrounding the instanton  then we have that the integrand in the surface ordered integral in \rf{charge} behaves as
\be
\(\alpha +\kappa\, \beta\) F_{\mu\nu}\,\frac{dx^{\mu}}{d\sigma}\frac{dx^{\nu}}{d\tau}\rightarrow \frac{1}{r^2} \qquad\qquad {\rm as} \qquad\qquad r\rightarrow \infty
\lab{instantonvanish}
\ee
where $r$ is the radius of $S^2_{\infty}$. Therefore, the operator \rf{charge} is unity, i.e. $Q_{S^2_{\infty}}=\one$, and that unity comes from the exponentiation of the trivial element of the Lie algebra. So, the one-instanton solution have indeed vanishing charges associated to \rf{charge} . 

Let us now evaluated the charges associated to the operator \rf{chargeselfdual}. Without any loss of generality let us take the circle $S_{\infty}^{1,(t)}$ of infinite radius to lie on the plane $x^{1}\,x^{2}$, at some constant values of $x^3$ and $x^4$. Due to the symmetries of the one-instanton solution the calculation on any other plane is very similar. We shall use polar coordinates on the plane $x^{1}=\rho\cos \sigma$, and $x^{2}=\rho\sin \sigma$, with $s^2=\rho^2+\(x^3\)^2+\(x^4\)^2$, and $\rho\rightarrow \infty$ , and where we have taken the polar angle $\sigma$ to be the same as the parameter which parameterizes the circle $S_{\infty}^{1,(t)}$. Therefore, using \rf{one-instanton}, the integrand of the path ordered integral in  \rf{chargeselfdual} becomes 
\be
A_{\mu}\,\frac{dx^{\mu}}{d\sigma}= -(2/e)\,\rho\( -\sigma_{1\nu}\sin\sigma+\sigma_{2\nu}\cos\sigma\)\frac{\(x^{\nu}-a^{\nu}\)}{\(x^{\mu}-a^{\mu}\)^2-\lambda^2}
\ee
 As $\rho\rightarrow \infty$, the only non-vanishing terms are those where $x^{\nu}$ is one of the coordinates of the plane, i.e. $x^{1}$ or $x^{2}$. Then $A_{\mu}\,\frac{dx^{\mu}}{d\sigma}\rightarrow (2/e)\, \sigma_{12}$, and so \rf{chargeselfdual} becomes 
\be
V\({\cal D}_{\infty}\) =P_1\,e^{-ie\oint_{S_{\infty}^{1,(t)}}d\sigma\, A_{\mu}\frac{dx^{\mu}}{d\sigma}}=
e^{-i\, 2\int_0^{2\pi}d\theta\, \sigma_{12}}=e^{-i\,4\,\pi\, T_3}
\lab{oneinstantonflux}
 \ee
 where ${\cal D}_{\infty}$ is the infinity disk with border $S_{\infty}^{1,(t)}$ on the plane $x^1\,x^2$. Adapting  the interpretation given below \rf{constantoperatoridentity}, to the euclidean case at hand, such operator should correspond to the (euclidean) magnetic flux $\Phi$ through  ${\cal D}_{\infty}$, i.e. $V\({\cal D}_{\infty}\)=e^{-i\,e\,\Phi\({\cal D}_{\infty}\)}$. If one takes finite dimensional representations of $SU(2)$, the eigenvalues of $T_3$ are integers or half integers and so $V\({\cal D}_{\infty}\)=\one$, which is compatible with the fact the connection $A_{\mu}$ for the one-instanton, is flat in the limit $s\rightarrow \infty$. However, that fact also implies that the flux should be quantized as
 \be
\Phi\({\cal D}_{\infty}\)= \frac{2\,\pi\,n}{e}\qquad\qquad \qquad \qquad 
n=0,\pm 1,\pm 2, \ldots
\lab{quantizeinstanton}
\ee 
However, following the same reasoning, the charges coming from \rf{charge} should also be associated, in such self-dual case solution, to the (euclidean) magnetic flux through the closed sphere $S^2_{\infty}$. But as we have shown below \rf{instantonvanish} that flux must vanish. Therefore, the only compatible value of $n$ in \rf{quantizeinstanton} seems to be $n=0$.

Let us now consider the case of the two-instanton solution. A closed form for the regular (non-singular) form of that solution is not easy. However we need only its asymptotic  form to calculate the charges and that is provided by Giambiagi and Rothe \cite{bocha}. Consider a two-instanton regular solutions where the position four-vector of each instanton is given by $a^{\mu}_1$ and $a^{\mu}_2$. Then the asymptotic form of the connection is given by \cite{bocha}
\be
A_{\mu}\rightarrow \frac{4}{e\,a^2\,s^2}\,\left[ \(x\cdot a\)\sigma_{\mu\lambda}\,b_{\lambda} + b_{\mu}\, x_{\nu}\,\sigma_{\nu\lambda}\,b_{\lambda}\right] \qquad\qquad {\rm as} \qquad\qquad s\rightarrow \infty
\lab{twoinstantonamu}
\ee 
where $s^2= x_1^2+x_2^2+x_3^2+x_4^2$, $\sigma_{\mu\nu}$ is the same as in \rf{sigmamunudef},  $a_{\mu}$ is the difference between the two position four-vectors, and $b_{\mu}$ is the reflection of $a_{\mu}$ through the hyperplane perpendicular to $x_{\mu}$, i.e.
\be
a^{\mu}\equiv a^{\mu}_1- a^{\mu}_2\qquad\qquad \qquad \qquad\qquad
b_{\mu}\equiv a_{\mu}-2\,\frac{\(x\cdot a\)}{x^2}\, x_{\mu}
\ee
and so $b^2=a^2$. 

The leading term of the connection given in \rf{twoinstantonamu} is flat, and it falls as $1/s$ as $s\rightarrow \infty$. Therefore, the leading term of the field tensor which would fall as $1/s^2$ vanishes, and therefore $F_{\mu\nu}$ falls at least as $1/s^3$. Consequently, the integrand of the surface ordered integral in \rf{charge}, namely $F_{\mu\nu}\,\frac{dx^{\mu}}{d\sigma}\frac{dx^{\nu}}{d\tau}$, falls faster than $1/s$, and so it vanishes in the limit $s\rightarrow \infty$. We then conclude that, similarly to the one-instanton  case, the charges associated to the operator \rf{charge} vanish when evaluated on the two-instanton solution. 

We now evaluate the charges associated to the operator \rf{chargeselfdual} for the two-instanton solution.  Given an infinite plane (disk) ${\cal D}_{\infty}$ with border being the circle $S_{\infty}^{1,(t)}$ of infinite radius we can choose, without loss of generality, the axis $x^1$ and $x^2$ to lie on that plane. We then split the vector $a^{\mu}$ in its perpendicular and parallel parts w.r.t. to the plane, i.e. $a^{\mu}= a^{\mu}_{\perp}+a^{\mu}_{\parallel}$, and take the axis $x^1$ to lie along $a^{\mu}_{\parallel}$, and  the axis $x^3$ to lie along $a^{\mu}_{\perp}$. In addition we take polar coordinates on the plane $x^1\,x^2$, such that $x^1=\rho\cos\sigma$, and $x^2=\rho\,\sin \sigma$, with the polar angle $\sigma$ being the same as the parameter in \rf{chargeselfdual} parametrizing $S_{\infty}^{1,(t)}$. Then the integrand of the path ordered integral in \rf{chargeselfdual}  along the infinite circle $S_{\infty}^{1,(t)}$ for the connection \rf{twoinstantonamu} becomes ($\rho\rightarrow \infty$) 
\be
A_{\mu}\,\frac{dx^{\mu}}{d\sigma}\rightarrow \frac{4}{e}\left[\frac{\mid a_{\parallel}\mid^2}{a^2}\, T_3+\frac{{\mid a_{\parallel}\mid}\, {\mid a_{\perp}\mid}}{a^2}\,\left[\cos\(2\sigma\)\,T_1+\sin\(2\sigma\)\, T_2\right]\right]
\ee
We now perform a gauge transformation $A_{\mu}\rightarrow \(g_2\,g_1\)\,A_{\mu}\, \(g_2\,g_1\)^{-1}+\frac{i}{e}\,\partial_{\mu}\(g_2\,g_1\)\,\(g_2\,g_1\)^{-1}$, with $g_1=e^{i\,\,2\,\sigma\,T_3}$ and $g_2=e^{i\,\,2\,\varphi\,T_2}$. The angle $\varphi$ is defined as follows: since $a^2=\mid a_{\parallel}\mid^2+\mid a_{\perp}\mid^2$, we parametrize it as 
$\mid a_{\parallel}\mid= \mid a\mid\,\cos \varphi$, and $\mid a_{\perp}\mid=\mid a\mid\,\sin\varphi$, with $0\leq \varphi\leq \frac{\pi}{2}$. So, since the vector $a_{\mu}$ was chosen to lie on the plane $x^1\,x^3$,  $\varphi$ is the angle between   $a_{\mu}$ and the plane $x^1\,x^2$ measured along the plane $x^1\,x^3$. Under such a gauge transformation one gets that
\be
A_{\mu}\,\frac{dx^{\mu}}{d\sigma}\rightarrow A_{\mu}^{\prime}\,\frac{dx^{\mu}}{d\sigma}= 
\frac{2}{e}\, T_3
\ee
and so
\be
P_1\,e^{-ie\oint_{S_{\infty}^1}d\sigma\, A_{\mu}\frac{dx^{\mu}}{d\sigma}}\rightarrow 
g_1\(\sigma=2\,\pi\)^{-1}\, g_2^{-1}\,P_1\,e^{-ie\oint_{S_{\infty}^1}d\sigma\, A_{\mu}^{\prime}\frac{dx^{\mu}}{d\sigma}}\, g_2\, g_1\(\sigma=0\)
\ee
Therefore, the operator \rf{chargeselfdual}  becomes
\be
V\({\cal D}_{\infty}\)  
= P_1\,e^{-ie\oint_{S_{\infty}^1}d\sigma\, A_{\mu}\frac{dx^{\mu}}{d\sigma}}
= e^{-i\,4\,\pi\,T_3}\,e^{-i\,\,2\,\varphi\,T_2}\, e^{-i\,\,4\,\pi\,T_3}\, e^{i\,\,2\,\varphi\,T_2}
\lab{twoinstantonflux}
\ee
We can try to interpret that result in terms of the (euclidean) magnetic flux $\Phi$ through the infinite disk ${\cal D}_{\infty}$. Since we are dealing with a non-abelian gauge theory one should not expect a linear superposition of the fluxes of each instanton. We have seen in \rf{oneinstantonflux} that the exponentiated flux of a single instanton is   $e^{-i\,4\,\pi\,T_3}$. In addition, since $\varphi$ is the angle between the line passing through the centers of the instantons and the disc ${\cal D}_{\infty}$, the result \rf{twoinstantonflux} could give a hint on how the fluxes compose. That is certainly a point which deserves further study. Again, as in any  finite dimensional representation of $SU(2)$ we have that $e^{-i\,4\,\pi\,T_3}=\one$, and so $V\({\cal D}_{\infty}\)=\one$, which is compatible with the fact that $A_{\mu}$ is flat at the leading order we have performed the calculation.  Using the flux interpretation of the charges we can write $V\({\cal D}_{\infty}\)=e^{-i\,e\,\Phi_{2-inst.}\({\cal D}_{\infty}\)}$, and so the two-instanton flux $\Phi_{2-inst.}\({\cal D}_{\infty}\) $ is quantized as in \rf{quantizeinstanton}.

 \subsection{Merons}
 \label{sec:merons}
 
 Merons are singular euclidean solutions not self-dual with one-half unit of topological charge \cite{alfaro}. We shall work here with such solutions in the Coulomb gauge since it is more suitable for the evaluation of the charges and it also connects with monopole solutions. The solution for a one-meron located at the origin is given by \cite{alfaro,actor}  
  \be
 A_i=-\frac{1}{e}\,\varepsilon_{ijk}\frac{{\hat r}_j}{r}\,\(1-\frac{x_4}{\sqrt{x_4^2+r^2}}\)\, T_k
 \qquad\qquad \qquad\qquad
 A_4=0 
 \lab{onemeron}
 \ee
 with $r^2=x_1^2+x_2^2+x_3^2$,  $i,j,k=1,2,3$, and $T_k$ are the generators of the $SU(2)$ Lie algebra. Note that for $x_4=0$ the connection \rf{onemeron} coincides with that for the Wu-Yang monopole given in \rf{monopole}. In addition, it interpoles between two vacuum configurations, i.e. for $x_4\rightarrow \infty$ the connection \rf{onemeron} vanishes, and for $x_4\rightarrow -\infty$ it becomes of a pure gauge form $A_i= \frac{i}{e}\,\partial_i  g\,g^{-1}$, with $g=\exp\(i\,\pi\,{\hat r}\cdot T\)$.
 
 In order to evaluate the charges \rf{charge} we need the field tensor at infinity, which is given by
 \be
 F_{ij}\rightarrow \frac{1}{e}\,\varepsilon_{ijk}\, \frac{{\hat r}_k}{r^2}\; {\hat r}\cdot T
 \qquad\qquad\qquad
 F_{4i}\rightarrow \frac{1}{e}\,\varepsilon_{ijk}\frac{{\hat r}_j}{r^2}\,T_k
 \qquad\qquad\qquad r\rightarrow \infty
 \ee
 Note that when taking the limit $r\rightarrow \infty$ we have kept $x_4$ finite. The double limit $r\rightarrow \infty$ and $x_4\rightarrow \pm \infty$, is not well defined. The asymptotic form of the space components of the dual tensor is ($\varepsilon_{1234}=1$)
 \be
 {\widetilde F}_{ij}\rightarrow -\frac{1}{e}\,\frac{1}{r^2}\left[ {\hat r}_i\, T_j - {\hat r}_j\, T_i\right] \qquad\qquad\qquad r\rightarrow \infty
 \ee 
If we evaluate the operator \rf{charge} on a spatial two-sphere $S^2_{\infty}$ of infinite radius and centered at the origin, it turns out that ${\hat r}$ is perpendicular to  $S^2_{\infty}$ and the derivatives $\frac{dx^i}{d\sigma}$ and $\frac{dx^i}{d\tau}$, with $\sigma$ parametrizing the loops scanning the sphere and $\tau$ labeling them, are parallel to $S^2_{\infty}$. Therefore, we have that ${\widetilde F}_{ij}\, \frac{dx^i}{d\sigma}\,\frac{dx^j}{d\tau}=0$. Consequently, the calculation of the operator \rf{charge} for the one-meron solution is identical to that for the monopole (see calculation leading to \rf{qsmonopole}). So, we have that
\br
Q_S= P_2\,e^{i\,e\,\int_{S^2_{\infty}}d\sigma\,d\tau\,W^{-1}\,\left[\alpha\,F_{ij}+\beta\, {\widetilde F}_{ij}\right]W\,\frac{dx^i}{d\sigma}\,\frac{dx^j}{d\tau} }=e^{-i\,e\,\alpha\,\int_{S^2_{\infty}}d{\vec\Sigma}\cdot{\vec B}^R }
=e^{-i\,e\,\alpha\,G_R }
=e^{i\,4\,\pi\,\alpha\,T_R}
 \lab{qsonemeron}
\er
where we have introduced a (euclidean) magnetic field in a way similar to that in \rf{qsmonopole}, i.e. $B^R_i\equiv -\frac{1}{2}\,\varepsilon_{ijk}\, W^{-1}\,F_{jk}\,W= -\frac{1}{e}\, \frac{{\hat r}_i}{r^2}\, T_R$, with $T_R$ being the value of ${\hat r}\cdot T$ at the reference point $x_R$ used in the scanning of the sphere . Using the same arguments as in the case of the monopole we conclude that the magnetic charges $G_R$ are quantized as in \rf{quantizemagnetic}. 

We then conclude that the one-meron solution has a magnetic charge $G_R$ conserved in the euclidean time $x_4$, and it is quantized in units of $\frac{2\,\pi}{e}$. What it is not clear is what happens to that charge in the limit $x_4\rightarrow \pm \infty$, since as we have seen, the connection \rf{onemeron} becomes flat in that limit, and so the charge should disappear. One of the difficulties in answering that is the fact that the double limit $r\rightarrow \infty$ and $x_4\rightarrow \pm \infty$, of the connection \rf{onemeron} is not well defined.   

The evaluation of the charges \rf{constantoperatoridentity} for the one-meron solution is also identical to that of the monopole, leading to \rf{halflux}. Indeed, if we consider the circle $S_{\infty}^{1,(t)}$ of infinite radius to lie on spatial planes, then only the components $A_i$, $i=1,2,3$, of the connection matters. But in the limit $r\rightarrow \infty$ the connection \rf{onemeron} becomes identical to that for the monopole \rf{monopole}. Therefore, following the arguments leading to \rf{halflux} one gets that
\be
V^{\prime}_{x_R^{(t)}}\({\cal D}_{\infty}^{(t)}\)  = P_1\,e^{-ie\oint_{S_{\infty}^{1,(t)}}d\sigma\, A_{i}\frac{dx^{i}}{d\sigma}}=e^{-i\,2\,\pi\,T_3}
\lab{halfluxonemeron}
\ee
The eigenvalues of that operator are conserved in the euclidean time $x_4$, and their interpretation, given below \rf{halflux}, in terms of the magnetic flux through the surface which border is $S_{\infty}^{1,(t)}$ remains valid. Again, we do not know what happens to those charges in the limit $x_4\rightarrow \pm \infty$, for the same reasons given above in the case of the one-meron magnetic charge. 
 
 The two-meron solution in the Coulomb gauge corresponding to one meron siting at the position $x^{\mu}=a^{\mu}=\(0,0,0,a\)$ and the other at $x^{\mu}=b^{\mu}=\(0,0,0,b\)$ is given by \cite{alfaro,actor}
 \be
 A_i=-\frac{1}{e}\,\varepsilon_{ijk}\frac{{\hat r}_j}{r}\,\(1+
 \frac{r^2+\(x_4-a\)\(x_4-b\)}{\sqrt{\(x-a\)^2\(x-b\)^2}}\)\, T_k
 \qquad\qquad \qquad\qquad
 A_4=0 
 \lab{twomeron}
 \ee
Expanding it in powers of $\frac{1}{r}$ one gets that
\be
A_{i}=\frac{i}{e}\,\partial_i  g\,g^{-1} 
+\frac{1}{e}\,\frac{\(a-b\)^2}{2}\,\varepsilon_{ijk}\frac{{\hat r}_j}{r^3}\,T_k
+O\(\frac{1}{r^5}\)
\ee
 with $g=\exp\(i\,\pi\,{\hat r}\cdot T\)$, and so the leading term is of pure gauge form, i.e. it is flat. Therefore, we have that
 \be
 F_{ij}\sim O\(\frac{1}{r^4}\) \qquad\qquad \qquad 
 F_{0i}\sim O\(\frac{1}{r^5}\) 
 \ee
Consequently, the integrand in \rf{charge}, namely $(\alpha F_{ij}+\beta {\widetilde F}_{ij}) \,\frac{dx^{i}}{d\sigma}\frac{dx^{j}}{d\tau}$, behaves as $O\(\frac{1}{r^2}\)$ in the limit $r\rightarrow \infty$. Therefore, the charges associated to \rf{charge} vanish, i.e. $Q_S=\one$.

Note that in the limit $r\rightarrow \infty$ the spatial component of the connection \rf{twomeron} for the two-meron solution is twice that of the one-meron solution \rf{onemeron}. Therefore, the evaluation of the charges associated to the operator 
\rf{constantoperatoridentity} is very similar to that leading to \rf{halfluxonemeron} and gives
\be
V^{\prime}_{x_R^{(t)}}\({\cal D}_{\infty}^{(t)}\)  = P_1\,e^{-ie\oint_{S_{\infty}^{1,(t)}}d\sigma\, A_{i}\frac{dx^{i}}{d\sigma}}=e^{-i\,4\,\pi\,T_3}=\one
\lab{halfluxotwomeron}
\ee
where the last equality follows from the fact that the leading term of $A_i$ is flat. The interpretation for such conserved charges,  given below \rf{constantoperatoridentity}, holds true, i.e. they correspond to the magnetic flux through the surface which border is $S_{\infty}^{1,(t)}$, and such fluxes are also quantized. 
 
 The meron-antimeron solution in the Coulomb gauge, corresponding to a meron and an anti-meron located at $x^{\mu}=-a^{\mu}$ and $x^{\mu}=a^{\mu}$ respectively, with $a^{\mu}=\(0,0,0,a\)$,  is given by \cite{alfaro,actor} 
 \be
 A_i=-\frac{1}{e}\,\varepsilon_{ijk}\frac{{\hat r}_j}{r}\,\(1-
 \frac{x^2-a^2}{\sqrt{\(x+a\)^2\(x-a\)^2}}\)\, T_k
 \qquad\qquad \qquad\qquad
 A_4=0 
 \ee
Again expanding in powers of $\frac{1}{r}$ one gets
\be
A_{i}=-\frac{2\,a^2}{e}\,\varepsilon_{ijk}\frac{{\hat r}_j}{r^3}\,T_k
+O\(\frac{1}{r^5}\)
\ee 
Then, similarly to the two-meron case one has that $F_{ij}\sim O\(\frac{1}{r^4}\)$ and $F_{0i}\sim O\(\frac{1}{r^5}\)$, and so the charges coming from \rf{charge} are trivial, i.e. $Q_S=\one$. In addition, since the connection falls faster than $\frac{1}{r}$ the integrand in the operator \rf{constantoperatoridentity} vanishes, i.e. $P_1\,e^{-ie\oint_{S_{\infty}^{1,(t)}}d\sigma\, A_{i}\frac{dx^{i}}{d\sigma}}=\one$. The corresponding charges are also trivial in this case. 

\subsection{Summary of the charges}

As we have seen in the examples above the surface and path ordering are not necessary in the evaluation of the operators \rf{charge} and \rf{chargeselfdual} (or equivalently \rf{constantoperatoridentity}) which eigenvalues are the charges. Therefore, those operators can be written as an ordinary exponential of Lie algebra elements. The only exception is the charge for the two-instanton solution associated to the operator \rf{chargeselfdual} which involve a non-linear superposition of fluxes (see eq. \rf{twoinstantonflux}). Therefore, we shall write the operator \rf{charge}, defined on a volume inside a two-sphere  $S^2_{\infty}$ of infinite radius, as
\be
P_2 e^{ie\int_{S^2_{\infty}}d\tau d\sigma\, W^{-1}\, (\alpha F_{\mu\nu}+\beta {\widetilde F}_{\mu\nu}) \,W\,\frac{dx^{\mu}}{d\sigma}\frac{dx^{\nu}}{d\tau}}= 
e^{-i\,e\,\left[\alpha\,G_R +\beta\, K_R\right]}
\ee
where $G_R$ and $K_R$ are the magnetic and electric charges defined as
\be
 G_R=\int_{S^2_{\infty}}d{\vec\Sigma}\cdot{\vec B}^R 
 \qquad\qquad\qquad\qquad\qquad
 K_R=\int_{S^2_{\infty}}d{\vec\Sigma}\cdot{\vec E}^R
 \ee
 with $B^R_i\equiv -\frac{1}{2}\,\varepsilon_{ijk}\, W^{-1}\,F_{jk}\,W$ and $E_i^R=W^{-1}\,F_{0i}\,W$, being respectively the abelian magnetic and electric fields. 
 
Similarly we shall write the operators  \rf{chargeselfdual} and \rf{constantoperatoridentity} on a surface ${\cal D}_{\infty}$ with the border being a circle $S_{\infty}^1$ of infinite radius, as
\be
P_1\,e^{-ie\oint_{S_{\infty}^1}d\sigma\, A_{\mu}\frac{dx^{\mu}}{d\sigma}}=
e^{-i\,e\,\Phi\({\cal D}_{\infty}\)}
\ee
where $\Phi\({\cal D}_{\infty}\)$ is interpreted as the magnetic flux through the surface ${\cal D}_{\infty}$.

Note that the magnetic charge $G_R$  is quantized due to the integrable Bianchi identity \rf{nontrivial}. The electric  charge $K_R$ and the magnetic flux $\Phi\({\cal D}_{\infty}\)$ are quantized only if we evaluate the operators 
\rf{charge},  \rf{chargeselfdual} and \rf{constantoperatoridentity} on a finite dimensional representation of the $SU(2)$ gauge group, where the eigenvalues of the generator $T_3$ (or any other element conjugated to it) are integers or half-integers, which we shall denote by $\frac{n}{2}$, with $n$ integer. 

With those definitions we summarize the spectrum of charges of the solutions discussed above in the  Table 1.\\

\begin{table}[htb]

\centering
\begin{tabular}{|c|c|c|c|}
\hline
& $G_R$ & $K_R$ &  $\Phi\({\cal D}_{\infty}\)$\\
\hline
\hline
Monopole & $\frac{2\,\pi\,n}{e}$ & $0$ & $\frac{\pi\,n}{e}$\\
\hline
Dyon & $\frac{2\,\pi\,n}{e}$ & $\frac{2\,\pi\,\gamma\,n}{e}$ & $\frac{\pi\,n}{e}$\\
\hline
$1$-instanton & $0$ & $0$ & $\frac{2\,\pi\,n}{e}$\\
\hline
$2$-instanton & $0$ & $0$ & see \rf{twoinstantonflux}\\
\hline
$1$-meron & $\frac{2\,\pi\,n}{e}$ & $0$ & $\frac{\pi\,n}{e}$\\
\hline
$2$-meron & $0$ & $0$ & $\frac{2\,\pi\,n}{e}$\\
\hline
meron-antimeron & $0$ & $0$ & $0$\\
\hline
\end{tabular}
\caption{Eigenvalues of $G_R$, $K_R$ and  $\Phi\({\cal D}_{\infty}\)$, for $SU(2)$ solutions}
\end{table}

\appendix

\section{Regularization of Wilson lines}
\label{sec:wilson}
\setcounter{equation}{0}

\begin{figure}[ht]
\centering
   \includegraphics[width=1\textwidth]{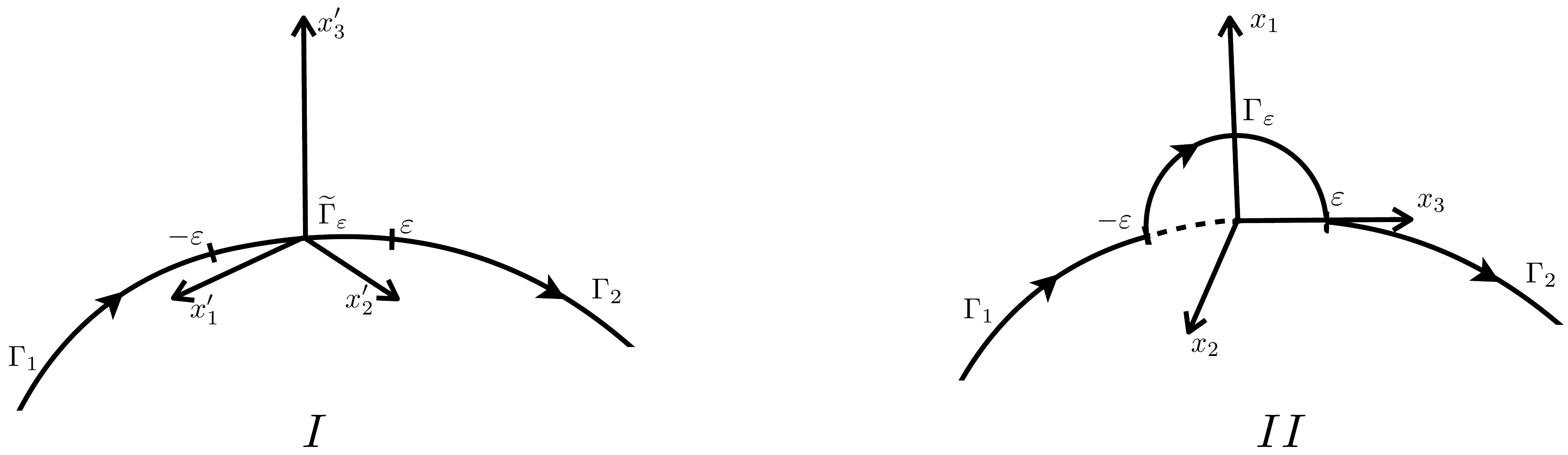}
   \caption{The regularization of the Wilson line operator is done by replacing the path that passes through to the origin by a path going around it.}
 \label{fig:regularize}
\end{figure}

In order to calculate the r.h.s. of the relations \rf{monopolevolumeformula} and \rf{dyonvolumeformula}, which give the conserved charges, as volume ordered integrals, for the Wu-Yang monopole and dyon solutions respectively, we have to evaluate the Wilson line operator $W$, defined by \rf{eqforw}, for the  connection  
\be
A_i =   -\frac{1}{e}\,\varepsilon_{ijk}\, \frac{x^j}{r^2}\, T_k
\lab{singularai}
\ee
which  has a singularity at the origin of the coordinate system. In the case of the dyon the time component of the connection is non-zero and also present a singularity at the origin. However, it does not play a role in the charge calculation since all the Wilson line operators are defined on space curves with no time component. We show in this appendix how the Wilson line operator can be regularized, when it is integrated along a purely spatial (no time) curve $\Gamma$  passing through the origin. In order to do that we shall split $\Gamma$ into three parts, $\Gamma=\Gamma_1\cup{\tilde \Gamma}_{\varepsilon}\cup\Gamma_2$, as shown in part I of the Figure \ref{fig:regularize}.

Therefore, the solution of \rf{eqforw} can be written as
\be
W=W\(\Gamma_2\)\,W\({\tilde \Gamma}_{\varepsilon}\)\,W\(\Gamma_1\)
\lab{partitionw}
\ee
The quantities $W\(\Gamma_1\)$ and $W\(\Gamma_2\)$ do not involve the singularity and so we should not worry about them. We have to evaluate $W\({\tilde \Gamma}_{\varepsilon}\)$ which pass through the origin. We shall take ${\tilde \Gamma}_{\varepsilon}$ infinitesimally small in such a way that we can approximate it by an infinitesimal straight line of length $2\,\varepsilon$ containing the origin in its middle  point. Note that the quantity $A_{i}\,\frac{d\,x^{i}}{d\,\sigma}$, appearing in \rf{eqforw}, is invariant under rotations, and so we can rotate the coordinate system in such a way that the $x^3$-axis lies parallel to ${\tilde \Gamma}_{\varepsilon}$ and in the direction of growing $\sigma$, i.e. in the sense of integration of  \rf{eqforw}, as shown in part II of the Figure \ref{fig:regularize}. Along such infinitesimal straight line ${\tilde \Gamma}_{\varepsilon}$, parametrized as $x^3=\sigma$, we have that  $A_{i}\,\frac{d\,x^{i}}{d\,\sigma}=-\frac{1}{e}\, \frac{1}{r^2}\left[ x^1\, T_2-x^2\,T_1\right]$. However, on ${\tilde \Gamma}_{\varepsilon}$ one has $x^1=x^2=0$, and so for $r\neq 0$ such expression vanishes. On the other hand, for $r=0$ it diverges, and so, we have a quite ill defined quantity. 

In order to regularize the Wilson line we shall replace  ${\tilde \Gamma}_{\varepsilon}$ by a semi-circle $\Gamma_{\varepsilon}$  of radius $\varepsilon$, with diameter being the previous straight line, and lying on the plane $x^1\, x^3$ as shown in part II of Figure \ref{fig:regularize}. We evaluate $W\(\Gamma_{\varepsilon}\)$ on such semi-circle and then take the limit $\varepsilon\rightarrow 0$. The points in $\Gamma_{\varepsilon}$ can be parameterized  as
\be
x^1=\varepsilon \,\sin \sigma \qquad\qquad x^2=0\qquad\qquad x^3=-\varepsilon \cos \sigma \qquad\qquad 0\leq \sigma \leq \pi
\ee
Therefore for all such points we have $r=\varepsilon$, and so from \rf{singularai} one has 
\br
A_{i}\,\frac{d\,x^{i}}{d\,\sigma}&=& \varepsilon\, \(A_1\,  \cos \sigma+A_3 \, \sin \sigma\)
= -\frac{1}{e}\, T_2
\er
We note that it does not depend upon $\varepsilon$ and  $\sigma$, and lies in the direction of just one generator of $SU(2)$. Therefore, the problem is abelian and the path ordering is not necessary. Then from \rf{eqforw} we have
\be
W\(\Gamma_{\varepsilon}\)=e^{i\,\pi\, T_2}
\lab{semicirclew1}
\ee
which we take as the regularized expression for $W\({\tilde \Gamma}_{\varepsilon}\)$.  Of course, we would obtain different results for different choices of curves going around the origin, specially non-planar curves. However, as we shown below, the evaluation of the  r.h.s. of the relations \rf{monopolevolumeformula} and \rf{dyonvolumeformula} is independent of such choices, and the regularization of those quantities is quite unique. 

Note that for the Wu-Yang monopole and dyon solutions one has that  
\be
F_{ij}=\frac{1}{e}\,\varepsilon_{ijk}\, \frac{{\hat r}_k}{r^2}\; {\hat r}\cdot T
\; ;  \qquad\qquad \qquad \quad
{\widetilde F}_{ij}=-\frac{\gamma}{e}\,\varepsilon_{ijk}\,\frac{{\hat r}_k}{r^2}\,{\hat r}\cdot T
\lab{wuyangtensors}
\ee
with $\gamma=0$ in the pure monopole case. In the evaluation of \rf{calcdejmonopole} and \rf{caljdyondef} we have to deal with the conjugated quantities $F_{ij}^W$ and ${\widetilde F}_{ij}^W$, and so essentially we have to worry about the quantity $W^{-1}\, {\hat r}\cdot T\,W$. 
 Our prescription is to scan the volume (the whole space) with closed surfaces based at $x_R$, and each of those surfaces are scanned with loops starting and ending at $x_R$. The origin lies on a given surface labeled by $\zeta_0$, and to just one loop, labeled by $\tau_0$, on that surface, and corresponding to the point labeled by $\sigma_0$ on that loop. For the surfaces corresponding to $\zeta < \zeta_0$ there are no problems in the integration since everything is regular. On each loop on those surfaces one can use  the relations \rf{rtcov} and \rf{rtconstant} to conclude that, along such loops, $W^{-1}\,{\hat r}\cdot T\,W$ is constant  and equal to $T_R$, i.e. the value of ${\hat r}\cdot T$ at the reference point $x_R$ where all loops are based. 

 Therefore, the commutators in \rf{calcdejmonopole} and \rf{caljdyondef}  vanish for $\zeta<\zeta_0$, since the conjugated tensors  $F_{ij}^W$ and ${\widetilde F}_{ij}^W$ all lie in the direction of $T_R$ on any point of any loop scanning the surfaces for $\zeta<\zeta_0$.  
On the surface for $\zeta=\zeta_0$ everything is fine until we reach the loop corresponding to $\tau=\tau_0$. In other words, the commutators in \rf{calcdejmonopole} and \rf{caljdyondef}  also vanish  for $\zeta=\zeta_0$ and $\tau<\tau_0$. Let us consider the loop corresponding to $\tau=\tau_0$. For $\sigma<\sigma_0$ we still have the vanishing of those commutators  since the singularity has not been touched yet. 
After crossing the singularity we have that the Wilson line $W$ becomes $W_2\, W\(\Gamma_{\varepsilon}\)\, W_1$ (see \rf{partitionw}), where $W_1$ is the result of the integration of \rf{eqforw} along $\Gamma_1$, i.e. the curve from the reference point $x_R$ up to the point marked $-\varepsilon$ on Figure \ref{fig:regularize}, along the loop corresponding to $\tau=\tau_0$, which passes through the origin. Similarly  $W_2$ is obtained by  integrating  \rf{eqforw} along $\Gamma_2$, i.e. the curve from  the point marked $\varepsilon$ on Figure \ref{fig:regularize}, up to some generic point beyond the origin along that  same loop. In addition, $W\(\Gamma_{\varepsilon}\)$ is the regularized expression, given in \rf{semicirclew1}, for the integration of \rf{eqforw} along ${\tilde \Gamma}_{\varepsilon}$.  

Along the curve $\Gamma_2$ the connection \rf{singularai} is regular, and so we can use  \rf{rtcov} and \rf{rtconstant} to  conclude that 
\be
W_2^{-1}\,{\hat r}\cdot T \,W_2=  \({\hat r}\cdot T\)_{\Gamma_2^0}
\ee
where $\({\hat r}\cdot T\)_{\Gamma_2^0}$ is the value of ${\hat r}\cdot T$ at the initial point of the curve $\Gamma_2$, which is the point marked $\varepsilon$ on Figure \ref{fig:regularize}. But since we have rotated the coordinate system such that the $x^3$-axis lies along ${\tilde \Gamma}_{\varepsilon}$, we have that $\({\hat r}\cdot T\)_{\Gamma_2^0}=T_3$. Now using  \rf{semicirclew1}, we have that 
\be
W^{-1}\(\Gamma_{\varepsilon}\)\,W_2^{-1}\,{\hat r}\cdot T \,W_2\, W\(\Gamma_{\varepsilon}\)=  e^{-i\,\pi\, T_2}\,T_3\, e^{i\,\pi\, T_2}= - T_3=
\({\hat r}\cdot T\)_{\Gamma_1^{\rm end}}
\lab{t3conjuga}
\ee
since $-T_3$ is the value of ${\hat r}\cdot T$ at the final point of the curve $\Gamma_1$, which is the point marked $-\varepsilon$ on Figure \ref{fig:regularize}. Along the curve $\Gamma_1$ the connection \rf{singularai} is regular, and so we can use  \rf{rtcov} and \rf{rtconstant} to get
\be
W_1^{-1}\,W^{-1}\(\Gamma_{\varepsilon}\)\,W_2^{-1}\,{\hat r}\cdot T \,W_2\, W\(\Gamma_{\varepsilon}\)\, W_1=W_1^{-1}\, \({\hat r}\cdot T\)_{\Gamma_1^{\rm end}}\, W_1= T_R
\ee
where $T_R$ is the value of $\({\hat r}\cdot T\)$ at the reference point $x_R$ which is the initial point of $\Gamma_1$. Therefore, the field tensor and its dual, given in \rf{wuyangtensors}, lie in the direction of $T_R$ when conjugated with $W_2\, W\(\Gamma_{\varepsilon}\)\, W_1$, and so the commutators in \rf{calcdejmonopole} and \rf{caljdyondef}  vanish when evaluated on the loop corresponding to $\tau=\tau_0$, i.e. the one passing through the singularity of \rf{singularai}. Of course, the quantities \rf{calcdejmonopole} and \rf{caljdyondef} will vanish on all loops scanning the surfaces for $\zeta >\zeta_0$, since the potential \rf{singularai} is not singular there, and the relations \rf{rtcov} and \rf{rtconstant} can be used to show that $W^{-1}\,{\hat r}\cdot T\,W=T_R$, for $W$ obtained by the integration of \rf{eqforw} on such loops.

Consequently  all the commutators in \rf{calcdejmonopole} and \rf{caljdyondef} vanish on any loop on the scanning of any surface on the scanning  of the volume. Since the Wu-Yang solutions have no sources we have ${\tilde J}_{123}=J_0=0$, and so ${\cal J}_{{\rm monopole}}$ and ${\cal J}_{{\rm dyon}}$ also vanishes. Therefore we conclude that the r.h.s. of  \rf{monopolevolumeformula} and \rf{dyonvolumeformula} are equal to unity, i.e. 
\be
 P_3e^{\int_{{\rm space}} d\zeta d\tau  V{\cal J}_{{\rm monopole}}V^{-1}}=\one \; ; \qquad\qquad \qquad
 P_3e^{\int_{{\rm space}} d\zeta d\tau  V{\cal J}_{{\rm dyon}}V^{-1}}=\one
 \lab{unityvolumeintwuyang}
\ee

We now come to the issue of the uniqueness of the regularization procedure. We have chosen to replace the segment ${\tilde \Gamma}_{\varepsilon}$ by the semi-cicle $\Gamma_{\varepsilon}$. Let us now analyze what happens to the quantity $W^{-1} \( \Gamma_{\varepsilon}\)\, \({\hat r}\cdot T\)_{\Gamma_1^{\rm end}}\, W\( \Gamma_{\varepsilon}\)=W^{-1} \( \Gamma_{\varepsilon}\)\, \(-T_3\)\, W\( \Gamma_{\varepsilon}\)$, when we make arbitrary infinitesimal variations on the semi-circle $\Gamma_{\varepsilon}$ keeping its end points fixed, i.e. the points marked $\varepsilon$ and $-\varepsilon$ on  Figure \ref{fig:regularize}.  We have
\br
\delta\left[W^{-1} \( \Gamma_{\varepsilon}\)\, \(-T_3\)\, 
W \( \Gamma_{\varepsilon}\) \right]&=& 
\sbr{W^{-1}\( \Gamma_{\varepsilon}\)\,\(-T_3\)\,W\( \Gamma_{\varepsilon}\)}{W^{-1} \( \Gamma_{\varepsilon}\)\,\delta W \( \Gamma_{\varepsilon}\) }
\nonumber\\
&=&
\sbr{T_3}{W^{-1} \( \Gamma_{\varepsilon}\)\,\delta W \( \Gamma_{\varepsilon}\) }
\er
where in the last equality we have used \rf{semicirclew1} and \rf{t3conjuga}. 
The variation of the Wilson line can be easily evaluated using for instances the techniques of section 2 of \cite{afs1}. When the end points of the curve $ \Gamma_{\varepsilon}$ are kept fixed one gets
\begin{equation}
W^{-1}( \Gamma_{\varepsilon} )\delta W( \Gamma_{\varepsilon} ) = 
\int_0^{\pi} d \sigma \, W^{-1} F_{ij} W {d x^{i}\over{d\sigma}}\delta x^{j}
\lab{var2pi}
\end{equation}
where $F_{ij}$ is the curvature, given in \rf{wuyangtensors}, of the connection \rf{singularai}, and where $W$ in the integrand in \rf{var2pi}, is obtained by integrating \rf{eqforw} along $ \Gamma_{\varepsilon}$, from its initial point at $\sigma=0$ to the point $\sigma=\sigma$ where the tensor  $F_{ij}$ is evaluated.  As long as the transformed curve does not pass through the singularity of the connection \rf{singularai}, the relations \rf{rtcov} and \rf{rtconstant} can be used to show that $W^{-1}\,{\hat r}\cdot T\,W=-T_3$, where $-T_3$ is the value of ${\hat r}\cdot T$ at the initial point of $ \Gamma_{\varepsilon}$. Therefore, the integrand in \rf{var2pi} always lies  in the direction of $T_3$, and so
\be
\delta\left[W^{-1} \( \Gamma_{\varepsilon}\)\, \(-T_3\)\, 
W \( \Gamma_{\varepsilon}\) \right]=0
\ee
Consequently any curve $\Gamma$, with the same end points as 
$ \Gamma_{\varepsilon}$, and that can be continuously deformed into 
$ \Gamma_{\varepsilon}$, satisfies $W^{-1} \( \Gamma_{\varepsilon}\)\, \(-T_3\)\, W \( \Gamma_{\varepsilon}\) =W^{-1} \( \Gamma\, \)\(-T_3\)\, W \( \Gamma\)=T_3$. That shows that our prescription for the regularization of the Wilson line is independent of the choice of the curve replacing the segment $ {\tilde \Gamma}_{\varepsilon}$.  

Note that the special role being played by $T_3$ is an artifact of our choice of the orientation of the coordinate axis w.r.t. the curve. 
Note in addition that our results do not imply that the Wilson line does not change. It is  just the conjugation of $T_3$ by the Wilson line that remains invariant. In the cases where the variation of the curve  lies on the same plane as 
$ \Gamma_{\varepsilon}$, then the Wilson line itself is invariant.  The reason is that  the r.h.s. of \rf{var2pi} measures the magnetic flux through the 
infinitesimal surface spanned by the variation, and since the magnetic field is radial it is parallel to such surface, and so  $\delta W( \Gamma_{\varepsilon} ) =0$ in such cases. 

\vspace{3cm}

{\bf Acknowledgements:} The authors are grateful to fruitful discussions with O. Alvarez, E. Castellano, P. Klimas, M.A.C. Kneipp, R. Koberle, J. S\'anchez-Guill\'en, N. Sawado and W. Zakrzewski. LAF is partially supported by CNPq, and GL is supported by a CNPq scholarship.

\end{document}